

%
%
%
%
%
%
%
\def\standardrisposta{s }\def\reducedrisposta{r }
\def\doublerisposta{d }\def\cartarisposta{e }\def\amsrisposta{y }
\newcount\ingrandimento \newcount\sinnota \newcount\dimnota
\newcount\unoduecol \newdimen\collhsize \newdimen\tothsize
\newdimen\fullhsize \newcount\controllorisposta \sinnota=1
\newskip\infralinea  \global\controllorisposta=0
\message{ ********    Welcome to PANDA macros (Plain TeX, AP, 1991)}
\message{ ******** }
\message{       You'll have to answer a few questions in lowercase.}
\message{>  Do you want it in double-page (d), reduced (r)}
\message{or standard format (s) ? }\read-1 to\risposta
\message{>  Do you want it in USA A4 (u) or EUROPEAN A4 (e)}
\message{paper size ? }\read-1 to\srisposta
\message{>  Do you have AMSFonts 2.0 (math) fonts (y/n) ? }
\read-1 to\arisposta
%
%
%
%
%
\ifx\risposta\standardrisposta \ingrandimento=1200
\message{>> This will come out UNREDUCED << }
\dimnota=2 \unoduecol=1 \global\controllorisposta=1 \fi
\ifx\risposta\reducedrisposta \ingrandimento=1095 \dimnota=1
\unoduecol=1  \global\controllorisposta=1
\message{>> This will come out REDUCED << } \fi
\ifx\risposta\doublerisposta \ingrandimento=1000 \dimnota=2
\unoduecol=2  \global\controllorisposta=1
\message{>> You must print this in LANDSCAPE orientation << } \fi
\ifnum\controllorisposta=0  \ingrandimento=1200
\message{>>> ERROR IN INPUT, I ASSUME STANDARD UNREDUCED FORMAT <<< }
\dimnota=2 \unoduecol=1 \fi
\magnification=\ingrandimento
%
%
%
%
\newdimen\eucolumnsize \newdimen\eudoublehsize \newdimen\eudoublevsize
\newdimen\uscolumnsize \newdimen\usdoublehsize \newdimen\usdoublevsize
\newdimen\eusinglehsize \newdimen\eusinglevsize \newdimen\ussinglehsize
\newskip\standardbaselineskip \newdimen\ussinglevsize
\newskip\reducedbaselineskip \newskip\doublebaselineskip
\eucolumnsize=12.0truecm    
\eudoublehsize=25.5truecm   
\eudoublevsize=6.5truein    
\uscolumnsize=4.4truein     
\usdoublehsize=9.4truein    
\usdoublevsize=6.8truein    
\eusinglehsize=6.5truein    
\eusinglevsize=24truecm     
\ussinglehsize=6.5truein    
\ussinglevsize=8.9truein    
\standardbaselineskip=16pt  
\reducedbaselineskip=14pt   
\doublebaselineskip=12pt    
%
%
\def\Portoffset{}
\def\Landoffset{}
%
%

%
\tolerance=10000
\parskip 0pt plus 2pt  
%
%
\ifx\risposta\standardrisposta \infralinea=\standardbaselineskip \fi
\ifx\risposta\reducedrisposta  \infralinea=\reducedbaselineskip \fi
\ifx\risposta\doublerisposta   \infralinea=\doublebaselineskip \fi
\ifnum\controllorisposta=0    \infralinea=\standardbaselineskip \fi
\ifx\risposta\doublerisposta   \Landoffset \else \Portoffset \fi
\ifx\risposta\doublerisposta \ifx\srisposta\cartarisposta
\tothsize=\eudoublehsize \collhsize=\eucolumnsize
\vsize=\eudoublevsize  \else  \tothsize=\usdoublehsize
\collhsize=\uscolumnsize \vsize=\usdoublevsize \fi \else
\ifx\srisposta\cartarisposta \tothsize=\eusinglehsize
\vsize=\eusinglevsize \else  \tothsize=\ussinglehsize
\vsize=\ussinglevsize \fi \collhsize=4.4truein \fi
%
%
%
%
\newcount\contaeuler \newcount\contacyrill \newcount\contaams
\font\ninerm=cmr9  \font\eightrm=cmr8  \font\sixrm=cmr6
\font\ninei=cmmi9  \font\eighti=cmmi8  \font\sixi=cmmi6
\font\ninesy=cmsy9  \font\eightsy=cmsy8  \font\sixsy=cmsy6
\font\ninebf=cmbx9  \font\eightbf=cmbx8  \font\sixbf=cmbx6
\font\ninett=cmtt9  \font\eighttt=cmtt8  \font\nineit=cmti9
\font\eightit=cmti8 \font\ninesl=cmsl9  \font\eightsl=cmsl8
\skewchar\ninei='177 \skewchar\eighti='177 \skewchar\sixi='177
\skewchar\ninesy='60 \skewchar\eightsy='60 \skewchar\sixsy='60
\hyphenchar\ninett=-1 \hyphenchar\eighttt=-1 \hyphenchar\tentt=-1
\def\bfmath{\cmmib}                 
\font\tencmmib=cmmib10  \newfam\cmmibfam  \skewchar\tencmmib='177
\font\tencmbsy=cmbsy10  \newfam\cmbsyfam  \skewchar\tencmbsy='60
\font\tencmcsc=cmcsc10  \newfam\cmcscfam
\ifnum\ingrandimento=1095

\font\capsone=cmcsc10 at 10.95pt 

\else

\font\capsone=cmcsc10 at 12pt 
\fi

\def\ttaarr{\bf}		
\def\ppaarr{\sl}		

%
%
%
\newfam\eufmfam \newfam\msamfam \newfam\msbmfam \newfam\eufbfam
\def\Loadeulerfonts{\global\contaeuler=1 \ifx\arisposta\amsrisposta
\font\teneufm=eufm10              
\font\eighteufm=eufm8 \font\nineeufm=eufm9 \font\sixeufm=eufm6
\font\seveneufm=eufm7  \font\fiveeufm=eufm5
\font\teneufb=eufb10              
\font\eighteufb=eufb8 \font\nineeufb=eufb9 \font\sixeufb=eufb6
\font\seveneufb=eufb7  \font\fiveeufb=eufb5
\font\teneurm=eurm10              
\font\eighteurm=eurm8 \font\nineeurm=eurm9
\font\teneurb=eurb10              
\font\eighteurb=eurb8 \font\nineeurb=eurb9
\font\teneusm=eusm10              
\font\eighteusm=eusm8 \font\nineeusm=eusm9
\font\teneusb=eusb10              
\font\eighteusb=eusb8 \font\nineeusb=eusb9
\else \def\eufm{\tt} \def\eufb{\tt} \def\eurm{\tt} \def\eurb{\tt}
\def\eusm{\tt} \def\eusb{\tt}    \fi}

\def\loadamsmath{\global\contaams=1 \ifx\arisposta\amsrisposta
\font\tenmsam=msam10 \font\ninemsam=msam9 \font\eightmsam=msam8
\font\sevenmsam=msam7 \font\sixmsam=msam6 \font\fivemsam=msam5
\font\tenmsbm=msbm10 \font\ninemsbm=msbm9 \font\eightmsbm=msbm8
\font\sevenmsbm=msbm7 \font\sixmsbm=msbm6 \font\fivemsbm=msbm5
\else \def\msbm{\bf} \fi \def\Bbb{\msbm} \def\symbl{\msam} \tenpoint}
\def\loadcyrill{\global\contacyrill=1 \ifx\arisposta\amsrisposta
\font\tenwncyr=wncyr10 \font\ninewncyr=wncyr9 \font\eightwncyr=wncyr8
\font\tenwncyb=wncyr10 \font\ninewncyb=wncyr9 \font\eightwncyb=wncyr8
\font\tenwncyi=wncyr10 \font\ninewncyi=wncyr9 \font\eightwncyi=wncyr8
\else \def\cyrill{\sl} \def\cyrilb{\sl} \def\cyrili{\sl} \fi\tenpoint}
\ifx\arisposta\amsrisposta
\font\sevenex=cmex7               
\font\eightex=cmex8  \font\nineex=cmex9
\font\ninecmmib=cmmib9   \font\eightcmmib=cmmib8
\font\sevencmmib=cmmib7 \font\sixcmmib=cmmib6
\font\fivecmmib=cmmib5   \skewchar\ninecmmib='177
\skewchar\eightcmmib='177  \skewchar\sevencmmib='177
\skewchar\sixcmmib='177   \skewchar\fivecmmib='177
\font\ninecmbsy=cmbsy9    \font\eightcmbsy=cmbsy8
\font\sevencmbsy=cmbsy7  \font\sixcmbsy=cmbsy6
\font\fivecmbsy=cmbsy5   \skewchar\ninecmbsy='60
\skewchar\eightcmbsy='60  \skewchar\sevencmbsy='60
\skewchar\sixcmbsy='60    \skewchar\fivecmbsy='60
\font\ninecmcsc=cmcsc9    \font\eightcmcsc=cmcsc8     \else
\def\cmmib{\fam\cmmibfam\tencmmib}\textfont\cmmibfam=\tencmmib
\scriptfont\cmmibfam=\tencmmib \scriptscriptfont\cmmibfam=\tencmmib
\def\cmbsy{\fam\cmbsyfam\tencmbsy} \textfont\cmbsyfam=\tencmbsy
\scriptfont\cmbsyfam=\tencmbsy \scriptscriptfont\cmbsyfam=\tencmbsy
\scriptfont\cmcscfam=\tencmcsc \scriptscriptfont\cmcscfam=\tencmcsc
\def\cmcsc{\fam\cmcscfam\tencmcsc} \textfont\cmcscfam=\tencmcsc \fi
\catcode`@=11
\newskip\ttglue
\gdef\tenpoint{\def\rm{\fam0\tenrm}
  \textfont0=\tenrm \scriptfont0=\sevenrm \scriptscriptfont0=\fiverm
  \textfont1=\teni \scriptfont1=\seveni \scriptscriptfont1=\fivei
  \textfont2=\tensy \scriptfont2=\sevensy \scriptscriptfont2=\fivesy
  \textfont3=\tenex \scriptfont3=\tenex \scriptscriptfont3=\tenex
  \def\mcal{\fam2 \tensy}  \def\mmit{\fam1 \teni}
  \textfont\itfam=\tenit \def\it{\fam\itfam\tenit}
  \textfont\slfam=\tensl \def\sl{\fam\slfam\tensl}
  \textfont\ttfam=\tentt \scriptfont\ttfam=\eighttt
  \scriptscriptfont\ttfam=\eighttt  \def\tt{\fam\ttfam\tentt}
  \textfont\bffam=\tenbf \scriptfont\bffam=\sevenbf
  \scriptscriptfont\bffam=\fivebf \def\bf{\fam\bffam\tenbf}
     \ifx\arisposta\amsrisposta    \ifnum\contaeuler=1
  \textfont\eufmfam=\teneufm \scriptfont\eufmfam=\seveneufm
  \scriptscriptfont\eufmfam=\fiveeufm \def\eufm{\fam\eufmfam\teneufm}
  \textfont\eufbfam=\teneufb \scriptfont\eufbfam=\seveneufb
  \scriptscriptfont\eufbfam=\fiveeufb \def\eufb{\fam\eufbfam\teneufb}
  \def\eurm{\teneurm} \def\eurb{\teneurb} \def\eusm{\teneusm}
  \def\eusb{\teneusb}    \fi    \ifnum\contaams=1
  \textfont\msamfam=\tenmsam \scriptfont\msamfam=\sevenmsam
  \scriptscriptfont\msamfam=\fivemsam \def\msam{\fam\msamfam\tenmsam}
  \textfont\msbmfam=\tenmsbm \scriptfont\msbmfam=\sevenmsbm
  \scriptscriptfont\msbmfam=\fivemsbm \def\msbm{\fam\msbmfam\tenmsbm}
     \fi      \ifnum\contacyrill=1     \def\cyrill{\tenwncyr}
  \def\cyrilb{\tenwncyb}  \def\cyrili{\tenwncyi}         \fi
  \textfont3=\tenex \scriptfont3=\sevenex \scriptscriptfont3=\sevenex
  \def\cmmib{\fam\cmmibfam\tencmmib} \scriptfont\cmmibfam=\sevencmmib
  \textfont\cmmibfam=\tencmmib  \scriptscriptfont\cmmibfam=\fivecmmib
  \def\cmbsy{\fam\cmbsyfam\tencmbsy} \scriptfont\cmbsyfam=\sevencmbsy
  \textfont\cmbsyfam=\tencmbsy  \scriptscriptfont\cmbsyfam=\fivecmbsy
  \def\cmcsc{\fam\cmcscfam\tencmcsc} \scriptfont\cmcscfam=\eightcmcsc
  \textfont\cmcscfam=\tencmcsc \scriptscriptfont\cmcscfam=\eightcmcsc
     \fi            \tt \ttglue=.5em plus.25em minus.15em
  \normalbaselineskip=12pt
  \setbox\strutbox=\hbox{\vrule height8.5pt depth3.5pt width0pt}
  \let\sc=\eightrm \let\big=\tenbig   \normalbaselines
  \baselineskip=\infralinea  \rm}
\gdef\ninepoint{\def\rm{\fam0\ninerm}
  \textfont0=\ninerm \scriptfont0=\sixrm \scriptscriptfont0=\fiverm
  \textfont1=\ninei \scriptfont1=\sixi \scriptscriptfont1=\fivei
  \textfont2=\ninesy \scriptfont2=\sixsy \scriptscriptfont2=\fivesy
  \textfont3=\tenex \scriptfont3=\tenex \scriptscriptfont3=\tenex
  \def\mcal{\fam2 \ninesy}  \def\mmit{\fam1 \ninei}
  \textfont\itfam=\nineit \def\it{\fam\itfam\nineit}
  \textfont\slfam=\ninesl \def\sl{\fam\slfam\ninesl}
  \textfont\ttfam=\ninett \scriptfont\ttfam=\eighttt
  \scriptscriptfont\ttfam=\eighttt \def\tt{\fam\ttfam\ninett}
  \textfont\bffam=\ninebf \scriptfont\bffam=\sixbf
  \scriptscriptfont\bffam=\fivebf \def\bf{\fam\bffam\ninebf}
     \ifx\arisposta\amsrisposta  \ifnum\contaeuler=1
  \textfont\eufmfam=\nineeufm \scriptfont\eufmfam=\sixeufm
  \scriptscriptfont\eufmfam=\fiveeufm \def\eufm{\fam\eufmfam\nineeufm}
  \textfont\eufbfam=\nineeufb \scriptfont\eufbfam=\sixeufb
  \scriptscriptfont\eufbfam=\fiveeufb \def\eufb{\fam\eufbfam\nineeufb}
  \def\eurm{\nineeurm} \def\eurb{\nineeurb} \def\eusm{\nineeusm}
  \def\eusb{\nineeusb}     \fi   \ifnum\contaams=1
  \textfont\msamfam=\ninemsam \scriptfont\msamfam=\sixmsam
  \scriptscriptfont\msamfam=\fivemsam \def\msam{\fam\msamfam\ninemsam}
  \textfont\msbmfam=\ninemsbm \scriptfont\msbmfam=\sixmsbm
  \scriptscriptfont\msbmfam=\fivemsbm \def\msbm{\fam\msbmfam\ninemsbm}
     \fi       \ifnum\contacyrill=1     \def\cyrill{\ninewncyr}
  \def\cyrilb{\ninewncyb}  \def\cyrili{\ninewncyi}         \fi
  \textfont3=\nineex \scriptfont3=\sevenex \scriptscriptfont3=\sevenex
  \def\cmmib{\fam\cmmibfam\ninecmmib}  \textfont\cmmibfam=\ninecmmib
  \scriptfont\cmmibfam=\sixcmmib \scriptscriptfont\cmmibfam=\fivecmmib
  \def\cmbsy{\fam\cmbsyfam\ninecmbsy}  \textfont\cmbsyfam=\ninecmbsy
  \scriptfont\cmbsyfam=\sixcmbsy \scriptscriptfont\cmbsyfam=\fivecmbsy
  \def\cmcsc{\fam\cmcscfam\ninecmcsc} \scriptfont\cmcscfam=\eightcmcsc
  \textfont\cmcscfam=\ninecmcsc \scriptscriptfont\cmcscfam=\eightcmcsc
     \fi            \tt \ttglue=.5em plus.25em minus.15em
  \normalbaselineskip=11pt
  \setbox\strutbox=\hbox{\vrule height8pt depth3pt width0pt}
  \let\sc=\sevenrm \let\big=\ninebig \normalbaselines\rm}
\gdef\eightpoint{\def\rm{\fam0\eightrm}
  \textfont0=\eightrm \scriptfont0=\sixrm \scriptscriptfont0=\fiverm
  \textfont1=\eighti \scriptfont1=\sixi \scriptscriptfont1=\fivei
  \textfont2=\eightsy \scriptfont2=\sixsy \scriptscriptfont2=\fivesy
  \textfont3=\tenex \scriptfont3=\tenex \scriptscriptfont3=\tenex
  \def\mcal{\fam2 \eightsy}  \def\mmit{\fam1 \eighti}
  \textfont\itfam=\eightit \def\it{\fam\itfam\eightit}
  \textfont\slfam=\eightsl \def\sl{\fam\slfam\eightsl}
  \textfont\ttfam=\eighttt \scriptfont\ttfam=\eighttt
  \scriptscriptfont\ttfam=\eighttt \def\tt{\fam\ttfam\eighttt}
  \textfont\bffam=\eightbf \scriptfont\bffam=\sixbf
  \scriptscriptfont\bffam=\fivebf \def\bf{\fam\bffam\eightbf}
     \ifx\arisposta\amsrisposta   \ifnum\contaeuler=1
  \textfont\eufmfam=\eighteufm \scriptfont\eufmfam=\sixeufm
  \scriptscriptfont\eufmfam=\fiveeufm \def\eufm{\fam\eufmfam\eighteufm}
  \textfont\eufbfam=\eighteufb \scriptfont\eufbfam=\sixeufb
  \scriptscriptfont\eufbfam=\fiveeufb \def\eufb{\fam\eufbfam\eighteufb}
  \def\eurm{\eighteurm} \def\eurb{\eighteurb} \def\eusm{\eighteusm}
  \def\eusb{\eighteusb}       \fi    \ifnum\contaams=1
  \textfont\msamfam=\eightmsam \scriptfont\msamfam=\sixmsam
  \scriptscriptfont\msamfam=\fivemsam \def\msam{\fam\msamfam\eightmsam}
  \textfont\msbmfam=\eightmsbm \scriptfont\msbmfam=\sixmsbm
  \scriptscriptfont\msbmfam=\fivemsbm \def\msbm{\fam\msbmfam\eightmsbm}
     \fi       \ifnum\contacyrill=1     \def\cyrill{\eightwncyr}
  \def\cyrilb{\eightwncyb}  \def\cyrili{\eightwncyi}         \fi
  \textfont3=\eightex \scriptfont3=\sevenex \scriptscriptfont3=\sevenex
  \def\cmmib{\fam\cmmibfam\eightcmmib}  \textfont\cmmibfam=\eightcmmib
  \scriptfont\cmmibfam=\sixcmmib \scriptscriptfont\cmmibfam=\fivecmmib
  \def\cmbsy{\fam\cmbsyfam\eightcmbsy}  \textfont\cmbsyfam=\eightcmbsy
  \scriptfont\cmbsyfam=\sixcmbsy \scriptscriptfont\cmbsyfam=\fivecmbsy
  \def\cmcsc{\fam\cmcscfam\eightcmcsc} \scriptfont\cmcscfam=\eightcmcsc
  \textfont\cmcscfam=\eightcmcsc \scriptscriptfont\cmcscfam=\eightcmcsc
     \fi             \tt \ttglue=.5em plus.25em minus.15em
  \normalbaselineskip=9pt
  \setbox\strutbox=\hbox{\vrule height7pt depth2pt width0pt}
  \let\sc=\sixrm \let\big=\eightbig \normalbaselines\rm}
\gdef\tenbig#1{{\hbox{$\left#1\vbox to8.5pt{}\right.\n@space$}}}
\gdef\ninebig#1{{\hbox{$\textfont0=\tenrm\textfont2=\tensy
   \left#1\vbox to7.25pt{}\right.\n@space$}}}
\gdef\eightbig#1{{\hbox{$\textfont0=\ninerm\textfont2=\ninesy
   \left#1\vbox to6.5pt{}\right.\n@space$}}}
\def\alternativefont#1#2{\ifx\arisposta\amsrisposta \relax \else
\xdef#1{#2} \fi}
\global\contaeuler=0 \global\contacyrill=0 \global\contaams=0
%
%
%
%
\newbox\fotlinebb \newbox\hedlinebb \newbox\leftcolumn
\gdef\makeheadline{\vbox to 0pt{\vskip-22.5pt
     \fullline{\vbox to8.5pt{}\the\headline}\vss}\nointerlineskip}
\gdef\makehedlinebb{\vbox to 0pt{\vskip-22.5pt
     \fullline{\vbox to8.5pt{}\copy\hedlinebb\hfil
     \line{\hfill\the\headline\hfill}}\vss} \nointerlineskip}
\gdef\makefootline{\baselineskip=24pt \fullline{\the\footline}}
\gdef\makefotlinebb{\baselineskip=24pt
    \fullline{\copy\fotlinebb\hfil\line{\hfill\the\footline\hfill}}}
\gdef\doubleformat{\shipout\vbox{\makehedlinebb
     \fullline{\box\leftcolumn\hfil\columnbox}\makefotlinebb}
     \advancepageno}

\gdef\columnbox{\leftline{\pagebody}}
\gdef\line#1{\hbox to\hsize{\hskip\leftskip#1\hskip\rightskip}}
\gdef\fullline#1{\hbox to\fullhsize{\hskip\leftskip{#1}%
\hskip\rightskip}}
\gdef\footnote#1{\let\@sf=\empty
         \ifhmode\edef\#sf{\spacefactor=\the\spacefactor}\/\fi
         #1\@sf\vfootnote{#1}}
\gdef\vfootnote#1{\insert\footins\bgroup
         \ifnum\dimnota=1  \eightpoint\fi
         \ifnum\dimnota=2  \ninepoint\fi
         \ifnum\dimnota=0  \tenpoint\fi
         \interlinepenalty=\interfootnotelinepenalty
         \splittopskip=\ht\strutbox
         \splitmaxdepth=\dp\strutbox \floatingpenalty=20000
         \leftskip=\oldssposta \rightskip=\olddsposta
         \spaceskip=0pt \xspaceskip=0pt
         \ifnum\sinnota=0   \textindent{#1}\fi
         \ifnum\sinnota=1   \item{#1}\fi
         \footstrut\futurelet\next\fo@t}
\gdef\fo@t{\ifcat\bgroup\noexpand\next \let\next\f@@t
             \else\let\next\f@t\fi \next}
\gdef\f@@t{\bgroup\aftergroup\@foot\let\next}
\gdef\f@t#1{#1\@foot} \gdef\@foot{\strut\egroup}
\gdef\footstrut{\vbox to\splittopskip{}}
\skip\footins=\bigskipamount
\count\footins=1000  \dimen\footins=8in
\catcode`@=12
\tenpoint      
\newskip\olddsposta \newskip\oldssposta
\global\oldssposta=\leftskip \global\olddsposta=\rightskip

\gdef\yespagenumbers{\footline={\hss\tenrm\folio\hss}}
\gdef\ciao{\par\vfill\supereject \ifnum\unoduecol=2
     \if R\lrcol  \headline={}\nopagenumbers\null\vfill\eject
     \fi\fi \end}

\ifnum\unoduecol=1 \hsize=\tothsize   \fullhsize=\tothsize \fi
\ifnum\unoduecol=2 \hsize=\collhsize  \fullhsize=\tothsize \fi
\global\let\lrcol=L
\ifnum\unoduecol=1 \output{\plainoutput}\fi
\ifnum\unoduecol=2 \output{\if L\lrcol
       \global\setbox\leftcolumn=\columnbox
       \global\setbox\fotlinebb=\line{\hfill\the\footline\hfill}
       \global\setbox\hedlinebb=\line{\hfill\the\headline\hfill}
       \advancepageno
      \global\let\lrcol=R \else \doubleformat \global\let\lrcol=L \fi
       \ifnum\outputpenalty>-20000 \else\dosupereject\fi}\fi
\def\ifdoublepage{\ifnum\unoduecol=2 }
\def\filldots{\leaders\hbox to 1em{\hss.\hss}\hfill}
\def\inquadrb#1 {\vbox {\hrule  \hbox{\vrule \vbox {\vskip .2cm
    \hbox {\ #1\ } \vskip .2cm } \vrule  }  \hrule} }

\def\newline{\hfil\break}
\def\jump{\vskip\baselineskip} \newskip\iinnffrr
\def\sjump{\iinnffrr=\baselineskip
          \divide\iinnffrr by 2 \vskip\iinnffrr}
\def\bjump{\vskip\baselineskip \vskip\baselineskip}
\newcount\nmbnota  \def\clearnmbnota{\global\nmbnota=0}
\def\note#1{\global\advance\nmbnota by 1
    \footnote{$^{\the\nmbnota}$}{#1}}  \clearnmbnota
\def\setnote#1{\expandafter\xdef\csname#1\endcsname{\the\nmbnota}}
\newcount\nbmfig  \def\clearnbmfig{\global\nbmfig=0}
\gdef\figure{\global\advance\nbmfig by 1
      {\rm fig. \the\nbmfig}}   \clearnbmfig
\def\setfig#1{\expandafter\xdef\csname#1\endcsname{fig. \the\nbmfig}}
 \def\endformula{\eqno\numero $$}
 \def\efr{\endformula}
\newcount\frmcount \def\clearfrmcount{\global\frmcount=0}
\def\numero{\global\advance\frmcount by 1   \ifnum\indappcount=0
  {\ifnum\cpcount <1 {\hbox{\rm (\the\frmcount )}}  \else
  {\hbox{\rm (\the\cpcount .\the\frmcount )}} \fi}  \else
  {\hbox{\rm (\applett .\the\frmcount )}} \fi}
\def\nameformula#1{\global\advance\frmcount by 1%
\ifnum\draftnum=0  {\ifnum\indappcount=0%
{\ifnum\cpcount<1\xdef\spzzttrra{(\the\frmcount )}%
\else\xdef\spzzttrra{(\the\cpcount .\the\frmcount )}\fi}%
\else\xdef\spzzttrra{(\applett .\the\frmcount )}\fi}%
\else\xdef\spzzttrra{(#1)}\fi%
\expandafter\xdef\csname#1\endcsname{\spzzttrra}
\eqno \ifnum\draftnum=0 {\ifnum\indappcount=0
  {\ifnum\cpcount <1 {\hbox{\rm (\the\frmcount )}}  \else
  {\hbox{\rm (\the\cpcount .\the\frmcount )}}\fi}   \else
  {\hbox{\rm (\applett .\the\frmcount )}} \fi} \else (#1) \fi $$}
\def\nfr{\nameformula}    
\def\nameali#1{\global\advance\frmcount by 1%
\ifnum\draftnum=0  {\ifnum\indappcount=0%
{\ifnum\cpcount<1\xdef\spzzttrra{(\the\frmcount )}%
\else\xdef\spzzttrra{(\the\cpcount .\the\frmcount )}\fi}%
\else\xdef\spzzttrra{(\applett .\the\frmcount )}\fi}%
\else\xdef\spzzttrra{(#1)}\fi%
\expandafter\xdef\csname#1\endcsname{\spzzttrra}
  \ifnum\draftnum=0  {\ifnum\indappcount=0
  {\ifnum\cpcount <1 {\hbox{\rm (\the\frmcount )}}  \else
  {\hbox{\rm (\the\cpcount .\the\frmcount )}}\fi}   \else
  {\hbox{\rm (\applett .\the\frmcount )}} \fi} \else (#1) \fi}
\clearfrmcount
\newcount\cpcount \def\clearcpcount{\global\cpcount=0}
\newcount\subcpcount \def\clearsubcpcount{\global\subcpcount=0}
\newcount\appcount \def\clearappcount{\global\appcount=0}
\newcount\indappcount \def\clearindappcount{\indappcount=0}
\newcount\sottoparcount 

\def\applett{\ifcase\appcount  \or {A}\or {B}\or {C}\or
{D}\or {E}\or {F}\or {G}\or {H}\or {I}\or {J}\or {K}\or {L}\or
{M}\or {N}\or {O}\or {P}\or {Q}\or {R}\or {S}\or {T}\or {U}\or
{V}\or {W}\or {X}\or {Y}\or {Z}\fi
             \ifnum\appcount<0
    \message{>>  ERROR: counter \appcount out of range <<}\fi
             \ifnum\appcount>26
   \message{>>  ERROR: counter \appcount out of range <<}\fi}
\clearappcount  \clearindappcount
\newcount\connttrre  \def\clearconnttrre{\global\connttrre=0}
\newcount\countref  \def\clearcountref{\global\countref=0}
\clearcountref
\def\chapter#1{\global\advance\cpcount by 1 \clearfrmcount
                 \goodbreak\null\vbox{\jump\nobreak
                 \clearsubcpcount\clearindappcount
                 \itemitem{\ttaarr\the\cpcount .\qquad}{\ttaarr #1}
                 \par\nobreak\jump\sjump}\nobreak}
\def\section#1{\global\advance\subcpcount by 1 \goodbreak\null
               \vbox{\sjump\nobreak\ifnum\indappcount=0
                 {\ifnum\cpcount=0 {\itemitem{\ppaarr
               .\the\subcpcount\quad\enskip\ }{\ppaarr #1}\par} \else
                 {\itemitem{\ppaarr\the\cpcount .\the\subcpcount\quad
                  \enskip\ }{\ppaarr #1} \par}  \fi}
                \else{\itemitem{\ppaarr\applett .\the\subcpcount\quad
                 \enskip\ }{\ppaarr #1}\par}\fi\nobreak\jump}\nobreak}
\clearsubcpcount
\def\appendix#1{\global\advance\appcount by 1 \clearfrmcount
                  \goodbreak\null\vbox{\jump\nobreak
                  \global\advance\indappcount by 1 \clearsubcpcount
                  \itemitem{\ }{\ttaarr #1}
                  \nobreak\jump\sjump}\nobreak}
\clearappcount \clearindappcount
\def\references{\goodbreak\null\vbox{\jump\nobreak
   \itemitem{}{\ttaarr References} \nobreak\jump\sjump}\nobreak}

\def\introduction{\clearindappcount\clearappcount\clearcpcount
                  \clearsubcpcount\goodbreak\null\vbox{\jump\nobreak
  \itemitem{}{\ttaarr Introduction} \nobreak\jump\sjump}\nobreak}
\clearcpcount\clearcountref
\def\acknowledgements{\goodbreak\null\vbox{\jump\nobreak
\itemitem{ }{\ttaarr Acknowledgements} \nobreak\jump\sjump}\nobreak}
\def\setchap#1{\ifnum\indappcount=0{\ifnum\subcpcount=0%
\xdef\spzzttrra{\the\cpcount}%
\else\xdef\spzzttrra{\the\cpcount .\the\subcpcount}\fi}
\else{\ifnum\subcpcount=0 \xdef\spzzttrra{\applett}%
\else\xdef\spzzttrra{\applett .\the\subcpcount}\fi}\fi
\expandafter\xdef\csname#1\endcsname{\spzzttrra}}
\newcount\draftnum \newcount\ppora   \newcount\ppminuti
\global\ppora=\time   \global\ppminuti=\time
\global\divide\ppora by 60  \draftnum=\ppora
\multiply\draftnum by 60    \global\advance\ppminuti by -\draftnum
\global\draftnum=0
\def\droggi{\number\day /\number\month /\number\year\ \the\ppora
:\the\ppminuti}

\global\draftnum=0
\def\draftcomment#1{\ifnum\draftnum=0 \relax \else {\ {\bf ***}\ #1\
{\bf ***}\ }\fi}

%
%
\catcode`@=11
\gdef\Ref#1{\expandafter\ifx\csname @rrxx@#1\endcsname\relax%
{\global\advance\countref by 1%
\ifnum\countref>200%
\message{>>> ERROR: maximum number of references exceeded <<<}%
\expandafter\xdef\csname @rrxx@#1\endcsname{0}\else%
\expandafter\xdef\csname @rrxx@#1\endcsname{\the\countref}\fi}\fi%
\ifnum\draftnum=0 \csname @rrxx@#1\endcsname \else#1\fi}
\gdef\beginref{\ifnum\draftnum=0  \gdef\Rref{\fairef}
\gdef\endref{\scriviref} \else\relax\fi}
\def\Reflab#1{[#1]}
\gdef\Rref#1#2{\item{\Reflab{#1}}{#2}}  \gdef\endref{\relax}
\newcount\conttemp
\gdef\fairef#1#2{\expandafter\ifx\csname @rrxx@#1\endcsname\relax
{\global\conttemp=0
\message{>>> ERROR: reference [#1] not defined <<<} } \else
{\global\conttemp=\csname @rrxx@#1\endcsname } \fi
\global\advance\conttemp by 50
\global\setbox\conttemp=\hbox{#2} }
\gdef\scriviref{\clearconnttrre\conttemp=50
\loop\ifnum\connttrre<\countref \advance\conttemp by 1
\advance\connttrre by 1
\item{\Reflab{\the\connttrre}}{\unhcopy\conttemp} \repeat}
\clearcountref \clearconnttrre
\catcode`@=12

\def\slashchar#1{\setbox0=\hbox{$#1$} \dimen0=\wd0
     \setbox1=\hbox{/} \dimen1=\wd1 \ifdim\dimen0>\dimen1
      \rlap{\hbox to \dimen0{\hfil/\hfil}} #1 \else
      \rlap{\hbox to \dimen1{\hfil$#1$\hfil}} / \fi}
\ifx\oldchi\undefined \let\oldchi=\chi
  \def\cchi{{\raise 1pt\hbox{$\oldchi$}}} \let\chi=\cchi \fi
\def\square{\hbox{{$\sqcup$}\llap{$\sqcap$}}}

\def\frac#1#2{{\textstyle{#1 \over #2}}}

\def\half{\ifinner {\scriptstyle {1 \over 2}}\else {1 \over 2} \fi}

\def\simge{\rlap{\raise 2pt \hbox{$>$}}{\lower 2pt \hbox{$\sim$}}}
\def\simle{\rlap{\raise 2pt \hbox{$<$}}{\lower 2pt \hbox{$\sim$}}}

\def\vbig#1#2{{\vbigd@men=#2\divide\vbigd@men by 2%
\hbox{$\left#1\vbox to \vbigd@men{}\right.\n@space$}}}

\null
%
%
%
%

\loadamsmath
\nopagenumbers{\baselineskip=12pt
\line{\hfill OUTP-92-03P}
\line{\hfill March, 1992}
\ifdoublepage \bjump\bjump\bjump\bjump\else\vfill\fi
\centerline{\capsone QUANTIZING SL(N) SOLITONS AND THE HECKE ALGEBRA}
\bjump\bjump
\centerline{\tencmcsc Timothy Hollowood\footnote{$^*$}
{holl\%dionysos.thphys@prg.oxford.ac.uk}}
\sjump
\centerline{\sl Theoretical Physics, 1 Keble Road,}
\centerline{\sl Oxford, OX1 3NP, U.K.}
\vfill
\ifnum\unoduecol=2 \eject\null\vfill\fi
\centerline{\capsone ABSTRACT}
\sjump
\noindent The problem of quantizing a class of two-dimensional
integrable quantum field theories is considered. The classical equations of the
theory are the complex $sl(n)$ affine Toda equations which admit
soliton solutions with real masses. The classical scattering theory of
the solitons is developed using Hirota's solution techniques.
A form for the soliton $S$-matrix is
proposed based on the constraints of $S$-matrix theory,
integrability and the
requirement that the semi-classical limit is consistent
with the semi-classical WKB quantization of the classical scattering theory.
The proposed $S$-matrix is an intertwiner of the quantum group
associated to $sl(n)$, where the deformation parameter is a function
of the coupling constant. It is further shown that the $S$-matrix
describes a non-unitary theory, which reflects the fact that the
classical Hamiltonian is complex. The spectrum of the theory is found
to consist of the basic solitons, scalar states (or breathers) and
excited (or `breathing') solitons. It is also noted that the
construction of the $S$-matrix is valid for any representation of the
Hecke algebra, allowing the definition of restricted $S$-matrices, in
which case the theory is unitary.
\sjump
\ifnum\unoduecol=2 \vfill\fi
\eject}
\yespagenumbers\pageno=1
\def\ba{\bar a}

\introduction

It is an outstanding problem to understand non-perturbative effects in
quantum field theory. In two-dimensions the situation promises to be
more tractable. Massless Euclidean
quantum field theories, which describe the critical scaling behaviour of
two-dimensional statistical systems, exhibit an infinite dimensional symmetry,
described by the group of, possibly extended, conformal transformations. In a
sense, the infinite problem of quantum field theory is reduced to a
finite problem involving the representation theory of an infinite
algebra [\Ref{bpz}].
In the space of massive two-dimensional
field theories, there is a class of theories that shares the
property of integrability with the conformal field theories.
In a massive integrable theory there exists an infinite number of commuting
conserved charges, and so there exists some transformation to action-angle
variables, and the theory is separable (at least classically).
For the relativistically invariant
classical integrable theories, it is interesting to
speculate as to the nature of the associated relativistic quantum
field theories. It is
thought that the property of integrability will generally survive
quantization and the resulting theory will be particularly simple
because its $S$-matrix will factorize [\Ref{factorize},\Ref{zam}].

For all their simplicity, it has still proved to be very difficult to
quantize the classical integrable theories.
There are a number of approaches, for
instance the quantum inverse scattering formalism [\Ref{qis}] or
the formalism of ref. [\Ref{leclair}]. Another approach, which has proved
particularly useful, might be called the `direct $S$-matrix approach'
[\Ref{zam}]. The success
of this latter approach, rests on the fact that the constraints of
$S$-matrix theory, along with the hitherto mentioned property of
factorizability, strongly constrain the allowed form of the
$S$-matrix, in fact to such an extent that it becomes possible, with a bit of
foresight, to conjecture a form for the $S$-matrix, up to possible
ambiguities of CDD type.

The $S$-matrix approach is particularly well adapted to quantizing
classical soliton theories because in such theories there is a direct
relationship between the semi-classical limit of the $S$-matrix and the
classical soliton scattering solutions [\Ref{semiclass}].
In a classical soliton theory the scattering of solitons, as described
by multi-soliton solutions to classical equations, can be
described in a very simple way. In the distant past, an $N$-soliton
solution has the form of $N$ well separated solitons of given
velocity. In the far future, the solution also consists of $N$
solitons with exactly the same distribution of velocities; this is due to
the existence of conserved higher spin integrals of motion.
If the scattering solution is analysed in
detail, then one finds that only the centres-of-mass of the solitons
change (consistent with the overall conservation of momentum).
A shift in the centre-of-mass of
a soliton can alternatively be thought of as a time-delay or advance.
The important point is that the scattering of the $N$
solitons can be thought of in terms of $\frac12N(N-1)$ elementary
pairwise scatterings,
each of which contributes to the shift in the centres-of-mass of the
participating solitons.

The picture in the quantum theory is thought to be very
similar in the sense that the scattering of $N$ solitons can also
be thought of in terms of pairwise scattering, with the property that
individual velocities are conserved (this is the factorizability
assumption [\Ref{factorize},\Ref{zam}]).
In the quantum theory there can be non-trivial mixings
due to mass degeneracies (such processes are seen in the classical
theory as solutions involving a complex time trajectory [\Ref{reflect}])
and the analogue of the time-delay is now
played by a non-trivial (momentum dependent)
phase factor. The semi-classical limit of the
phase factor is directly related to the classical time-delay.
So there is a very concrete connexion between the $S$-matrix and
classical scattering theory. The idea is to use the axioms of
$S$-matrix theory and the semi-classical limit in order to propose a
form for the $S$-matrix of a classical soliton scattering theory.

The known class of relativistic
integrable field theories includes the sine and
sinh-Gordon theories, the Toda field theories, the chiral models
and various fermion
models. In this work, we will consider the affine Toda theories associated to
the algebra $sl(n)$, which include the sine and sinh-Gordon
theories as special cases when $n=2$.

The sinh and sine-Gordon theories are almost completely understood, in
the sense that their complete spectra and $S$-matrices are known
[\Ref{zam},\Ref{korep}].
Since we shall argue that the more general Toda theories share many
features of the sine/sinh-Gordon theory it is worth discussing these
theories in some detail.
The equation of  motion for both theories may be written\note{
The notation used is not conventional, usually
$\beta\rightarrow\beta/\sqrt2$, however, it will
prove more useful for comparing with the $sl(n)$ generalization.}
$$
\square\phi=-{2m^2\over\beta}\sin
(\sqrt2\beta\phi).
\nfr{sglag}
The field $\phi(x,t)$ is a scalar, whilst $m$ and $\beta$ are coupling
constants. The
sinh-Gordon theory differs only in that $\beta$ is taken to be purely
imaginary.

The spectrum of the sinh-Gordon theory is particularly
simple, there is only a single scalar particle, which is identified
with $\phi$ the fundamental field of the theory.
The situation for the sine-Gordon theory is, in comparison,
much more complicated.
Classically the sine-Gordon theory admits soliton, or kink, solutions of
the form
$$
\phi(x,t)=\pm{2\sqrt2\over\beta}\tan^{-1}\left(e^{\sigma(x-vt-\xi)}
\right),
\nfr{sgkink}
where $v$, $\sigma$ and $\xi$ are constants, and
$\sigma^2(1-v^2)=4m^2$. The two solutions in \sgkink\ represent the
soliton and anti-soliton. Multiple soliton solutions also exist which
describe the scattering of solitons. There are other classical
solutions, known as the breather or doublet solutions, which have the
interpretation of a soliton and anti-soliton oscillating about a
fixed centre with some period. In a remarkable series of papers by
various authors [\Ref{korep},\Ref{wkb}],
the complete spectrum in the quantum  theory has been
determined. One has to distinguish two regions; the {\it attractive
regime\/} which, in our conventions, is
$0\leq\beta^2\leq 2\pi$ and the {\it repulsive regime\/},
$2\pi\leq\beta^2\leq4\pi$ (the language refers to the remarkable
connexion of the
sine-Gordon theory with the massive Thirring model [\Ref{coleman}]).
The theory is not
well-defined when $\beta^2>4\pi$. In the attractive regime, the
spectrum consists of a soliton and anti-soliton of mass
$$
\hat M={8m\over\beta^2}-{2m\over\pi}.
\nfr{sgsolmass}
The first term is the classical soliton mass, whilst the second is the
first quantum  correction. There are arguments to suggest that
\sgsolmass\ is actually exact to all order in $\beta^2$ [\Ref{wkb}].
In addition to the
solitons there is series of soliton anti-soliton bound states that
arise from quantizing the breather solutions, yielding a discrete mass spectrum
$$
\hat m(k)=2\hat M\sin\left({k\gamma\over8}\right),\ \ \ \ k=1,2,
\ldots,<{4\pi\over\gamma},
\nfr{massbr}
where
$$
\gamma={\beta^2\over1-\beta^2/4\pi}.
\nfr{defgam}
Remarkably, the ground state of the
breather system is identified with the fundamental particle of the
sine-Gordon Lagrangian. Indeed, $\hat m(1)=2m+O(\beta^2)$, where the
terms $O(\beta^2)$ agree, order by order,
with the loop corrections to the mass of the
fundamental particle. This theory exhibits the phenomenon of {\it
nuclear democracy\/}; the solitons can be though of as coherent
excitations of the fundamental particle, which itself can be though of
as the bound state of a soliton and anti-soliton. The full $S$-matrix
for the combined breather/soliton system was found in [\Ref{zam}].

In the repulsive regime the situation is not so straightforward
[\Ref{korep}]. The
vacuum becomes unstable and the spectrum is changed. New scalar states
appear which cannot be thought of as bound states of the solitons.

The sine-Gordon theory is also remarkable in that it can be
reformulated as the theory of a massive interacting fermion, namely
the massive Thirring model. In this picture the solitons are
represented by a Dirac spinor field and the breather states are then
fermion bound states [\Ref{coleman}].

We now turn our attention to the $sl(n)$ generalizations of the
sine/sinh Gordon theories. These are the affine $sl(n)$ Toda field theories.
The equation of motion of the theory is
$$
\square\phi=-{m^2\over \tilde\beta}
\sum_{j=1}^n\alpha_je^{\tilde\beta\alpha_j\cdot\phi}.
\nfr{lag}
The field $\phi(x,t)$ is an
$n-1\ (={\rm rank}\ sl(n))$-dimensional vector. The inner products are taken
with respect to the Killing form of $sl(n)$ restricted to the Cartan
subalgebra. The $\alpha_j$'s, for $j=1,\ldots,n-1$ are the simple roots
of $sl(n)$; $\alpha_n$ is the extended root (minus the highest
root)\note{For convenience we shall think of the labels on the
$\alpha_j$'s as being defined modulo $n$, so that
$\alpha_j\equiv\alpha_{j+n}$}.
 The fact that the extended root is included in the sum,
distinguishes the {\it affine\/} theories from the {\it non-affine\/} ones.
In the affine theories the $\alpha_j$'s are linearly dependent:
$$
\sum_{j=1}^n\alpha_j=0.
\efr
Notice that the sinh-Gordon theory is recovered by choosing
the algebra to be $sl(2)$.

For the moment we consider the theories which generalize the
sinh-Gordon theory, so $\tilde\beta$ is real. In the weak coupling
limit we expect the theory to include $n-1$ scalars with masses
$$
\hat m_a=m_a\left(1+{\tilde\beta^2\over4n}{\rm cot}{\pi\over n}+\cdots\right),
\quad a=1,2,\ldots,n-1,
\nfr{partmass}
where the first term is the `classical mass':
$$m_a=2m\sin\left({\pi a\over n}\right)\quad j=1,2,\ldots,n-1.
\nfr{classmass}
An $S$-matrix for the scalars was conjectured in [\Ref{arin}], which
is consistent with
the Feynman diagram expansion in the limit of weak coupling. The
$S$-matrix only has poles on the `physical strip' corresponding to the
exchange of the scalar particles, and so the spectrum is complete.
This just generalizes the situation for the sinh-Gordon theory.

Our interest is in soliton theories for which we need to
generalize the sine-Gordon theory. To do this we can consider the Toda
equations when $\tilde\beta$ is purely imaginary, so
$\tilde\beta=i\beta$. For $n>2$ this means that the equations are now
complex. Classically,
there exists a well-defined set of classical soliton solutions with
real masses. Rather than take the usual route to quantization we
shall proceed directly to an $S$ matrix, showing that our proposal is
consistent, in the semi-classical limit, with
the classical scattering theory.
Since the Hamiltonian of the theory is a
complex quantity (for $n>2$), we expect that the theory is non-unitary; indeed
the issue of unitarity can be addressed directly at the level of the
$S$-matrix.

Our proposal for the $S$-matrix involves the intertwiner of the
$sl(n)$ quantum group, and hence the Hecke algebra. In fact the
deformation parameter of the quantum group is the coupling constant
for the solitons, such that in the weak coupling limit of the solitons
the quantum group reduces the universal enveloping algebra of the Lie
algebra $sl(n)$ and the Hecke algebra reduces to the symmetric group,
which in physical terms means that the $S$-matrix is just a
permutation and hence describes a non-interacting theory. In crude
terms the $S$-matrix has a quantum group symmetry, as well as certain
momentum dependent symmetries which lead to new integrals of motion.
We also point out that the $S$-matrix can be constructed for any
representation of the Hecke algebra.

The plan of the paper is as follows. In \S1, we review the construction
of the classical soliton theory [\Ref{mysol}].
\S2 provides an introduction to the $sl(n)$
quantum group and the Hecke algebras, as well as proving a few results
which are needed in the following sections. A factorizable $S$-matrix
is constructed in \S3, which in \S4 is proposed as the soliton
$S$-matrix of the complex $sl(n)$ Toda theories. Various properties of
the $S$-matrix are investigated, for instance we show that the spectrum
consists of additional scalar states as well as excited solitons. In
addition we show, as expected,
that the $S$-matrix describes a non-unitary quantum field theory.
In \S5 we show that the $S$-matrix is
consistent with the semi-classical quantization of the classical
scattering theory, via a WKB approximation. Finally, some comments are made
in \S6.

\chapter{The Classical Solitons}

The soliton solutions of the complex $sl(n)$ Toda theory
were constructed in [\Ref{mysol}]. Here, we recall some of the details in order
to motivate the construction of the $S$-matrix in following sections.

The equation of motion of the complex $sl(n)$ affine Toda field theory is
$$
\square\phi=-{m^2\over i\beta}\sum_{j=1}^n\alpha_je^{i\beta\alpha_j
\cdot\phi}.
\nfr{eqnmotion}
This equation is integrable in the sense that there exists an infinite
number of Poisson-commuting conserved charges (see for example
[\Ref{olivet}]), one of which is
interpreted as the energy. This is the integral of the density
$$
H=\frac12\left((\partial_t\phi)^2+(\partial_x\phi)^2\right)
-{m^2\over\beta^2}\sum_{j=1}^n\left(e^{i\beta\alpha_j\cdot\phi}-1
\right).
\nfr{ham}

Notice that the field is periodic with respect to the weight
lattice, $\Lambda^\star$, of $sl(n)$. More precisely
$$
\phi\sim\phi+{2\pi\over\beta}w,\quad \forall w\in\Lambda^\star.
\efr
The constant field configurations $\phi=(2\pi/\beta)w$, $\forall
w\in\Lambda^\star$, have zero energy. There exist kink, or soliton, solutions
which interpolate between these configurations as $x$ goes from
$-\infty$ and $\infty$. One can associate a topological charge
$$
t={\beta\over2\pi}\int_{-\infty}^\infty dx\,\partial_x\phi\in
\Lambda^\star,
\nfr{topcharge}
to each solution.

Explicit expressions for the soliton solutions were
found in [\Ref{mysol}] using the Hirota formalism [\Ref{hirota}].
The idea is as follows: first one performs a change of variables
$$
\phi=-{1\over i\beta}\sum_{j=1}^n\alpha_j\log\tau_j.
\nfr{taujerk}
The equation of motion \eqnmotion\ is equivalent to
$$
\ddot\tau_j\tau_j-\dot\tau_j^2-\tau_j^{\prime\prime}\tau_j+\tau_j^{
\prime2}=m^2(\tau_{j-1}\tau_{j+1}-\tau_j^2),\quad j=1,\ldots,n,
\nfr{eqtau}
where, as for the roots, $\tau_j\equiv\tau_{j+n}$.
\eqtau\ is an equation of `Hirota
bilinear type' [\Ref{hirota}]. The characteristic polynomial in this case is
$$
{\cal F}(\sigma,\lambda,a)=\sigma^2-\lambda^2-4m^2\sin^2\left({\pi a
\over n}\right),
\nfr{poly}
where $\sigma$ and $\lambda$ are continuous variables and
$a\in\{1,2,\ldots,n-1\}$. It is useful to introduce
$v=\lambda/\sigma$, which is the velocity of a soliton.
The $N$-soliton solution is
then written in terms of functions $\Phi^{(p)}$, $p=1,2,\ldots,N$,
one for each soliton, and $\gamma^{(pq)}$ associated to each
soliton pair, $1\leq p<q\leq N$, where
$$
\Phi^{(p)}_j(x,t)=\sigma_px-\lambda_pt+{2\pi i\over n}a_pj+\xi_p,\quad
j=1,\ldots,n,
\nfr{phis}
subject to the constraint
$$
{\cal F}(\sigma_p,\lambda_p,a_p)=0.
\nfr{disprel}
In the above $\xi_p$
is a constant, whose real part represents an arbitrary shift in the
centre-of-mass, and whose imaginary part will determine the
topological charge of the soliton.
The `interaction function' is given by
$$
\exp \gamma^{(pq)}=-{{\cal
F}(\sigma_p-\sigma_q,\lambda_p-\lambda_q,a_p-a_q)
\over{\cal F}(\sigma_p+\sigma_q,\lambda_p+\lambda_q,a_p+a_q)},
\nfr{intfunct}
where ${\cal F}$ is the characteristic polynomial in \poly.
Another useful way to write the interaction function is
$$
\exp\gamma^{(pq)}(\theta)={\sin\left({\theta\over2i}+{\pi(a_p-a_q)\over2n}
\right)\sin\left({\theta\over2i}-{\pi(a_p-a_q)\over2n}\right)
\over\sin\left({\theta\over2i}+{\pi(a_p+a_q)\over2n}\right)
\sin\left({\theta\over2i}-{\pi(a_p+a_q)\over2n}\right)},
\nfr{intconst}
where $\theta=\theta_p-\theta_q$ is the rapidity difference of
the two solitons\note{The rapidity is related to the velocity by
$v={\rm tanh}\theta$, {\it i.e.\/} ${\rm tanh}\theta=\lambda/\sigma$.}.
The general $N$ soliton solution is then
$$
\tau_j(x,t)=\sum_{\mu_1=0}^1\cdots\sum_{\mu_N=0}^1\exp\left(\sum_{p=1}^N
\mu_p\Phi_j^{(p)}+\sum_{1\leq p<q\leq N}\mu_p\mu_q\gamma^{(pq)}\right).
\nfr{nsolitonsol}

Let us now analyse the one-soliton solution in more detail. The explicit
expression for the solution is
$$
\phi(x,t)=-{1\over
i\beta}\sum_{j=1}^n\alpha_j\log\left\{1+\exp\left(\sigma
(x-vt)+\xi+{2\pi ia\over n}j\right)\right\},
\nfr{onesol}
with
$$
\sigma^2(1-v^2)=4m^2\sin^2\left({\pi a\over n}\right).
\nfr{disponesol}
Clearly the solution represents a kink, whose centre is at $vt-\sigma^{-1}{\rm
Re}(\xi)$, moving with velocity $v$ and with characteristic size
$\sigma^{-1}$.
\disponesol\ simply expresses the relativistic invariance of the
theory: the the faster a soliton goes the
narrower it becomes.

We now find the topological charge of the {\it elementary\/}
soliton, that is the soliton with $a=1$. Assuming $\sigma>0$,
as $x\rightarrow-\infty$
$\phi\rightarrow0$. The limit as $x\rightarrow\infty$ is more
subtle and depends upon the choice for the imaginary  part of the
constant $\xi$. One obtains $n$ different values, for instance
with the choice
$$
{\rm Im}(\xi)=\cases{{2\pi (m-1)\over n}&$m=1,2,\ldots,n\quad n\ {\rm
odd}$\cr {\pi (2m-1)\over n}\quad&$m=1,2,\ldots,n\quad n\ {\rm even}$.\cr}
\efr
At this point it is convenient to introduce the weights of the $n$
dimensional representation of $sl(n)$, $e_j$ (again $e_{j+n}\equiv
e_j$). These
vectors have the property that $\sum_{j=1}^ne_j=0$, and the simple
roots, along with $\alpha_n$, are given by $\alpha_j=e_j-e_{j+1}$.
By using the $n$ possibilities for $\xi$
one finds that the topological charges of the elementary soliton fill out
the set of weights of the the $n$ dimensional representation.

Similarly one can show that the solitons for other values of $a$ have
topological charges in the set of weights of the $a^{\rm th}$ {\it
fundamental\/} representation of $sl(n)$\note{The fundamental
representations are the irreducible representations whose
highest weights are dual to the simple roots.}. In addition to the
soliton solutions one can also construct breather solutions. They are
obtained by considering the two soliton solution, for solitons of equal
mass, {\it i.e.\/} either two solitons of type $a$ or one of type $a$
and one of type $n-a$. In the centre-of-mass frame the breather
solution is obtained by taking $\lambda_1=-\lambda_2=i\omega$, for
$\omega\in{\Bbb R}$. The solution is interpreted as the two solitons
oscillating about a common centre. The new feature of the more general
theories is that breather solutions can have non-zero topological
charge, and so there are `breathing soliton' solutions.

One can easily calculate the masses of the soliton solutions from the
Hamiltonian \ham. The resulting masses only depend on the particular
fundamental representation that the topological charge of the soliton
lies in, one finds [\Ref{mysol}]
$$
M_a={4mn\over\beta^2}\sin\left({\pi a\over n}\right),\quad
a=1,\ldots,n-1.
\nfr{classsolmass}
Notice that these masses are proportional to the masses of the Toda
particles of the real coupling constant theory \classmass. It seems rather
miraculous that the resulting expressions for the masses are real
considering that the Hamiltonian is in general complex. However, this
can be traced to the fact that the soliton solutions satisfy a reality
condition of the form $\phi^\star=-M\phi$, where $M$ is an orthogonal
transformation with $M^2=1$, which acts as a ${\Bbb Z}_2$
permutation on the roots $\alpha_j$, $j=1,\ldots,n$.
The presence of the involution $M$ implies that energy of
a soliton is real, since
$H(\phi,\beta)^\star=H(-M\phi,-\beta)=H(\phi,\beta)$ due to invariance
of the Hamiltonian under permutation of the roots.
Some issues
concerning the reality of the complex Toda field theories are addressed
in ref. [\Ref{jon}]. The fact that the
masses are proportional to the masses of the fundamental Toda
particles is explained in ref. [\Ref{mysol}], via a relationship between the
bootstrap equations of the scattering theory of the fundamental
particles and a purely classical analogue of the bootstrap equations
for the soliton theory.

For the case when $n=2$ the doublet of soliton solutions of \onesol\ reduce to
the soliton and anti-soliton solutions of the sine-Gordon theory in \sgkink.

The multi-soliton solutions describe the scattering of the single
solitons. For example, consider an $N$ soliton solution with
$v_1>v_2\cdots>v_N$.
Initially, {\it i.e.\/} as $t\rightarrow-\infty$, the
solution is approaches a set of isolated solitons, in the order
$1,2,\ldots,N$ as $x$ goes from $-\infty$ to $\infty$; the position of the
$p^{\rm th}$ soliton at time $t$ being
$$
v_pt-{1\over\sigma_p}\left(\sum_{q=1}^{p-1}\gamma^{(pq)}+{\rm
Re}(\xi_p)\right).
\efr
Finally, {\it i.e.\/} as $t\rightarrow\infty$, the solution also
approaches that of $N$ isolated solitons moving with the {\it same\/}
set of velocities as those in the intial state, but now with order
$N,N-1,\ldots,1$ as $x$ goes from $-\infty$ to $\infty$, the position of the
$p^{\rm th}$ soliton being
$$
v_pt-{1\over\sigma_p}\left(\sum_{q=p+1}^N\gamma^{(pq)}+{\rm Re}(\xi_p)
\right).
\efr
If the solution is analysed in detail the following picture emerges.
Each soliton retains its integrity except when near another soliton,
within a distance $\sim m^{-1}$. In this interaction region the
solution is complicated, however, the solitons emerge with the same
velocities and the only effect of the interaction is to shift the
centre-of-mass of each soliton (with the overall centre-of-mass being
preserved). So an $N$ soliton scattering solution can be analysed in
terms of $\frac12 N(N-1)$ two-soliton scatterings. This is illustrated
in figure 1, where the circles indicate the regions where the solution
is not approximated by a set of isolated solitons; these are the
`interaction' regions.

\midinsert
\centerline{Figure 1. A multi-soliton scattering process.}
\endinsert

The shift in the
centre-of-mass of $q^{\rm th}$ soliton as it `interacts' with the
$p^{\rm th}$ soliton is
$-\gamma^{(pq)}/\sigma_q$. Another way to interpret this is to say that the
$q^{\rm th}$ soliton on interacting with the $p^{\rm th}$ soliton will
experience at time-delay given by
$-\gamma^{(pq)}/(\sigma_qv_q)$. This shift in the center-of-mass is
illustrated in figure 2.

\midinsert
\centerline{Figure 2. Two soliton scattering}
\endinsert

The question that we now address is: is there a factorizable
quantum soliton $S$-matrix whose semi-classical limit yields the
time-delays of the classical scattering theory described above? In our
search for the $S$-matrix we are guided by the situation for the
sine-Gordon soliton $S$-matrix. It is known that the $S$-matrix in
this case is, up to a scalar factor, the intertwiner of the $sl(2)$
quantum group. So it appears that we should consider the $sl(n)$ quantum groups
in order to construct the more general $S$-matrices. In the next
section we consider some relevant facts about the $sl(n)$ quantum group
and its commutant, the Hecke algebra. In \S3 and \S4 we go on to
propose a form for the factorizable $S$-matrix describing the quantum solitons.

\chapter{The $sl(n)$-Quantum Group and Hecke Algebra}

We have seen that classically each soliton mass eigenstate is
associated with a fundamental representation of $sl(n)$, in the sense
that the allowed topological charge is a weight of such a representation.
In the quantum theory we expect that the asymptotic state representing a
soliton to carry two quantum numbers, the velocity (or
rapidity) and the topological charge. Therefore associated to each
external state is a vector in one of the fundamental modules of $sl(n)$.
We denote the module corresponding to
the fundamental representation having Dynkin labels `one' over the $a^{\rm
th}$ simple root, and zero elsewhere, as $V_a$.
The soliton with mass $M_a$ is then associated to the module $V_a$.

Since the theory is integrable we assume that the
$S$-matrix is factorizable. This means that the individual momenta of each
external state is separately conserved, so there is no particle
creation and all the elements of the $S$-matrix can be deduced from the
two-body process [\Ref{zam}]. Notice that this is analogous to the
classical scattering theory of the solitons discussed in the last
section. In particular, a quantum scattering process may be analysed
in terms of pairwise scattering, so the multi-soliton $S$-matrix can
be constructed in terms of the two-body $S$-matrix.
{}From a Lie algebraic point of view, the two-body $S$-matrix
must act as the intertwiner:
$$
S^{a,b}:\ \ V_{a}\otimes V_{b}\mapsto V_{b}\otimes V_{a}.
\nfr{inter}
Notice that if there
were no interaction then the $S$-matrix would simply permute the
vector spaces.
This gives a clue about how to construct a non-trivial
$S$-matrix, since there exist natural generalizations of the
permutation groups known as the Hecke algebras, which reduce to the
former in some limit.

In the rest of this section we introduce the Hecke
algebras and the associated quantum groups. Our approach
follows that of M. Jimbo in refs. [\Ref{jimbo}].

\def\rc{\check R}
\def\ot{\otimes}
\def\sc{S}

Central to the subject is the Yang-Baxter Equation
(YBE) which can be written as an identity in
End$(V\otimes V\otimes V)$, for some vector space $V$:
$$(\rc(x)\ot I)(I\ot\rc(xy))(\rc(y)\ot I)
=(I\ot\rc(y))(\rc(xy)\ot I)(I\ot\rc(x)),
\nfr{AA}
where $\rc(x)\in{\rm End}(V\ot V)$, $I$ is the identity in End$(V)$,
and $x\in{\Bbb C}^\star$ is the {\it spectral parameter\/}.
We are interested in
the solutions for which $\rc(x)$ is trignometric in
$u=\log x$, and when $V\simeq{\Bbb C}^n$ is the $n$ dimensional
module of the Lie algebra $sl(n)$. The solution depends upon
an additional parameter $q$, the {\it deformation parameter\/},
which for the rest of this section we take to be {\it generic} (not equal
to a root of unity).

Trignometric solutions of the YBE are naturally described by
a {\it quantum group\/}; in this case the $q$-deformation of the
universal enveloping algebra of $g=sl(n)$, denoted $U_q(sl(n))$. The
parameter $q$ sets the degree of deformation and as $q\rightarrow1$,
$U_q(g)\rightarrow U(g)$. $U_q(g)$ is endowed with the structure of a
Hopf algebra. That is an algebra homomorphism
$$\Delta^{(m)}:\ U_q\longrightarrow U_q^{\ot m}.
\efr
$\Delta^{(m)}$ defines the action of $U_q(g)$ on tensor products of the
$n$ dimensional representation $\varrho:\,U_q(g)\rightarrow{\rm End}(V)$.
For generic $q$ (not equal to a root of unity)
the irreducible representations of $U_q(g)$ are in one-to-one
correspondence with those of $g$. The trignometric solution of the YBE
commutes with $(\varrho\times\varrho)
(\Delta^{(2)}(x))$, $x\in U_q(g)$, and so $\rc(x)$ lies in the
commutant of $\Delta^{(2)}(U_q(g))$.

In general, the commutant of $\Delta^{(m)}(U_q(g))$ is equal to the Hecke
algebra ${\cal H}_m$, which can be thought of as a deformation of the
symmetric group on $m$ objects, ${\cal S}_m$. A representation of
${\cal H}_m$ is generated by
$\{T_a,\ a=1,\ldots,m-1\}$, with $T_a\in{\rm End}(V^{\ot\,m})$,
subject to the following relations
$$\eqalign{(T_a-q^{-1})(T_a+q)&=0\cr T_aT_{a+1}T_a&=T_{a+1}T_aT_{a+1}\cr
[T_a,T_b]&=0\ \ \ |a-b|\geq2.\cr}
\nfr{AB}
In the limit $q\rightarrow1$ we recover the relations for the symmetric
group ${\cal S}_m$,
with $T_a\rightarrow\sigma_a$ being the generator which permutes
the $a^{\rm th}$ and $(a+1)^{\rm th}$ space in the tensor product.
In general, we
can label elements $T_w\in{\cal H}_m$ with elements $w\in{\cal S}_m$,
such that $T_{ww^\prime}=T_wT_{w^\prime}$ (if
$l(ww^\prime)=l(w)+l(w^\prime)$, where $l(w)$ is the {\it length\/} of
$w\in{\cal S}_m$).

It can be readily verified that the basic $\rc$ matrix can be expressed as
$$\rc(x)=xT^{-1}-x^{-1}T,
\nfr{AC}
which satisfies the YBE by virtue of the relations of the Hecke
algebra ${\cal H}_3$. To make the above discussion more explicit we
introduce a basis $\{{\bfmath e}_i,i=1,\ldots,n\}$ for $V$. From
ref. [\Ref{jimbo}] we have
$$T{\bfmath e}_i\otimes{\bfmath e}_j=\cases{q^{-1}{\bfmath
e}_i\otimes{\bfmath e}_i&$i=j$\cr (q^{-1}-q){\bfmath
e}_i\otimes{\bfmath e}_j+{\bfmath e}_j\otimes{\bfmath e}_i\ \
&$i>j$\cr {\bfmath e}_j\otimes{\bfmath e}_i &$i<j$.\cr}
\nfr{YYY}
The representation
constructed above actually satisfies an additional constraint, over
and above the Hecke algebra relations, this is the generalized Temperley-Lieb
condition, which is the vanishing of the deformed full
anti-symmetrizer $s_{n+1}^-$ defined below.

Just as one can generate higher irreducible
representations of $sl(n)$ by considering
the action of projection operators, formed from elements of the
symmetric group ${\cal S}_m$, encoded by the Young
Tableaux, on the $m$-fold tensor product of the basic
representation, so one can form higher irreducible representations of
$U_q(sl(n))$ by considering the same process with ${\cal S}_m$
replaced by ${\cal
H}_m$\note{For generic $q$, not equal to a root of unity.}.
In particular, we will be interested in the analogues of the {\it
fundamental\/} representations $\varrho_a$ of $U_q(sl(n))$, $a=1
,\ldots,n-1$, with
$\varrho_a:\ U_q(sl(n))\rightarrow\,{\rm End}(V_a)$
(where $\varrho_1\equiv\varrho$
and $V_1\equiv V$). $V_a$ can be projected out of the $a$-fold tensor
product $V_1^{\otimes\, a}$ with the Hecke algebra analogue of the
full anti-symmetrizer:
$$
V_a\simeq s_a^-\left(V^{\otimes\,a}\right).
\efr
An expression for the Hecke algebra
analogue of the full symmetrizer $s^+_m$ and anti-symmetrizer $s^-_m$
has been given in ref. [\Ref{jimbo}]
$$s_m^\pm={1\over[m]!}\sum_{w\in{\cal S}_m}(\pm)^{l(w)}q^{\pm(m(m-1)/2
-l(w))}T_w,
\nfr{AD}
where $[m]!=[m][m-1]\cdots\,[1]$ and $[m]=(q^m-q^{-m})/(q-q^{-1
})$. $s_m^\pm$ are projection operators, so that $(s_m^{\pm})^2
=s_m^\pm$. For example
$$s_2^+={1\over[2]}(q+T),\ \ \ s_2^-={1\over[2]}(q^{-1}-T),$$
and so in terms of these projection operators
$$\rc(x)=(xq-x^{-1}q^{-1})s_2^++(x^{-1}q-xq^{-1})s_2^-.
\nfr{AE}

Up till now, we have only considered solutions of the YBE for
$\rc(x)\in{\rm End}(V_1\ot V_1)$. One can also look for solutions
to the more general YBE
$$\eqalign{(\rc^{U_2,U_3}(x)\ot I_{U_1})&(I_{U_2}\ot\rc^{U_1,U_3
}(xy))(\rc^{U_1,U_2}(y)\ot
I_{U_3})\cr =&(I_{U_3}\ot\rc^{U_1,U_2}(y))(\rc^{U_1,U_3}(xy)\ot
I_{U_2})(I_{U_1}\ot\rc^{U_2,U_3}(x)),\cr}
\nfr{NAL}
where $\rc^{U_i,U_j}(x)=\sigma\cdot R^{U_i,U_j}(x)$, with
$$R^{U_i,U_j}(x)\in{\rm End}(U_i\ot U_j).
\nfr{AF}
Here, $\sigma$ is the permutation $\sigma(w\ot u)=u\ot w$ for $w\in U_i$,
$u\in U_j$ and the $U_i$ are arbitrary representations of $U_q(g)$.
Notice that both the left and right--hand sides of \NAL\ map $U_1\ot U_2\ot
U_3$ to $U_3\ot U_2\ot U_1$.

Since higher representations can be formed by taking tensor products of
the basic representation, the higher $\rc^{W,U}$--matrices can be built out
of the basic $\rc$--matrix, via the {\it fusion procedure\/}.
We briefly describe this process  following refs. [\Ref{jimbo}].

If we know $\rc^{U^{\prime\prime},U}(x)$ and $\rc^{U^{
\prime\prime},U^\prime}(x)$ then we can write down a new solution of the
YBE:
$$\rc^{U^{\prime\prime},U^\prime\ot
U}(x)=(I\ot\rc^{U^{\prime\prime},U}
(xy_1))(\rc^{U^{\prime\prime},U^\prime}(xy_2)\ot
I).
\nfr{AH}
With an appropriate choice of $y_1,y_2$ we can restrict the above to give
a new solution to the YBE $\rc^{U^{\prime\prime},W}(x)$, where $W$ is in the
decomposition of the tensor product $U^\prime\ot U$. The choice of $y_1
,y_2$ is determined by the requirement that
$$W\simeq \rc^{U^\prime,U}(y_2/y_1)(U^\prime\ot U)\subset U\ot U^\prime,
\nfr{AI}
{\it i.e.\/} is a proper subspace of $U\ot U^\prime$. A
simple application of the YBE then shows that
$\rc^{U^{\prime\prime},W}(x)$ so defined, is indeed a homomorphism from
$U^{\prime\prime}\ot W$ to $W\ot U^{\prime\prime}$.
So for example, notice that since $\rc((-q)^{-1})\propto s^-_2$ we have
$$V_2\simeq \rc((-q)^{-1})(V_1\ot V_1),$$
so we can find $\rc^{1,2}(x)$ acting on $V_1\ot V_2\simeq (I\ot s_
2^-)(V_1\ot V_1\ot V_1)$:
$$\rc^{1,2}(x)\equiv\rc^{V_1,V_2}(x)=(I\ot\rc(xy_1))
(\rc(xy_1(-q)^{-1})\ot I),
\nfr{AJ}
where $y_1$ is arbitrary. In a similar way one finds
$$\rc^{2,1}(x)\equiv\rc^{V_2,V_1}(x)=
(\rc(xy_1)\ot I)(I\ot\rc(xy_1(-q)^{-1})),
\nfr{AK}
acting on $V_2\ot V_1\simeq (s_2^-\ot I)(V_1\ot V_1\ot V_1)$.
By repeating the fusion procedure we can find $\rc^{a,b}(x)$, for $a,b=1
,\ldots,n-1$, acting in
$V_a\ot V_b\simeq (s_a^-\ot s_b^-)(V_1^{\ot\,(a+b)})$, a subspace of the
$a+b$-fold tensor product of $V_1$.

In general
[\Ref{jimbo}], $\rc^{a,b}(x)$ has the
following spectral decomposition
$$\rc^{a,b}(x)\equiv\rc^{V_a,V_b}(x)=\sum_{U}\rho_U(x)P_U,
\nfr{AG}
where $P_U$ is the orthogonal projector, relative to a $U_q(sl(n))$
invariant scalar product, onto the irreducible representation
$U$ in the tensor product $V_a\ot V_b\subset V_1^{\ot\,(a+b)}$, and
$\rho_U(x)$ is a scalar
function of $x$. Notice that \AE\ is of the form \AG, because $V_1\ot V_1=V_2
\oplus W$, since
$s_2^-$ is the projector onto the fundamental representation $V_2$,
and $W\simeq s_2^+(V_1\ot V_1)$ is the analogue of the symmetric tensor.
For generic $q$, this implies that between fundamental representations:
$$\rc^{a,b}(x)\equiv\rc^{V_a,
V_b}(x)=\rho_{a,b}(x)s_{a+b}^-+\cdots,$$
where the dots represent non-fundamental representations which appear in
the tensor product $V_a\ot V_b$, and the full anti-symmetrizer
$s_{a+b}^-$ is the projector
onto the unique fundamental representation $V_{a+b}$, which appears in
the tensor product, (and $a+b$ is taken modulo $n$).

Notice that at each stage of the fusion procedure
we are free to choose an overall shift in $x$, as
indicated by the presence  of $y_1$ in \AJ\ and \AK. We claim that the
following application of the fusion procedure defines a
solution to the YBE acting in the reducible module $V_1\oplus
V_2\oplus\cdots\oplus V_{n-1}$,
the sum of the fundamental representations, by recursion from the basic
solution in equation \AC:
$$\eqalign{\rc^{a,b+c}(x)&=(I\ot\rc^{a,b}(x(-q)^{c/2}))(\rc^{a,c}(x(-q)^{-b/2})
\ot I)\cr
\rc^{b+c,a}(x)&=(\rc^{b,a}(x(-q)^{c/2})\ot I)(I\ot\rc^{c,a}
(x(-q)^{-b/2})).\cr}
\nfr{AL}
At each stage we have set the overall multiplicative shift in $x$.
The first equation in \AL\ is to be understood as being
restricted to $V_a\ot V_{b+c}\simeq(s_a^-\ot s_{b+c}^-)V_1^{\ot
(a+b+c)}$ and the second to
$V_{b+c}\ot V_a\simeq(s_{b+c}^-\ot s_a^-)V_1^{\ot(a+b+c)}$. In
the above we assume that $a,b,c<n$ and $b+c<n$. The
YBE will be satisfied, by virtue of the fusion procedure, if \AI\ is
true, {\it i.e.\/}
$$\rc^{c,b}\left((-q)^{-(c+b)/2}\right)\propto s_{c+b}^-,
\nfr{AN}
since $s^-_{b+c}$ is the appropriate projection operator.
Since we have not managed to find an economical proof of \AN, the
details have been relegated to appendix A.

In the following section we will need to know the positions of any zeros
that
$\rc^{a,b}(x)$ might have. Using the fusion equations we have
$$\rc^{2,1}((-q)^{1/2})=(\rc(-q)\ot I)(I\ot\rc(1))(s_2^-\ot I).$$
However, $\rc(1)=(q-q^{-1})I\ot I$ and $\rc(-q)=-(q^2-q^{-2})s_2^+$,
therefore
$$\rc^{2,1}((-q)^{1/2})=-(q^2-q^{-2})(q-q^{-1})(s_2^+\ot I)(s_2^-\ot I)=0.$$
More generally,
by using the fusion equations, one can easily show that
$\rc^{a,b}(x)$, for $a\geq b$,
has zeros at $x=(-q)^{-(a+b-2i-2j)/2}$, for $i=1,2,\ldots,a-1$ and
$j=1,2,\ldots,b$.

Finally we note that
using properties of the Hecke algebra alone one can show
$$\rc(x)\rc(x^{-1})=(q^{-1}x-qx^{-1})(q^{-1}x^{-1}-qx)I\ot I,$$
and so by using \AL\ we have
$$\rc^{a,b}(x)\rc^{b,a}(x^{-1})=\prod_{i,j=1}^{a,b}g(x(-q)^{-(a+b-2i-2j+2)/
2})I\ot I,\nfr{AO}
where
$$g(x)=(q^{-1}x-qx^{-1})(q^{-1}x^{-1}-qx).$$

This concludes our discussion of the quantum group $U_q(sl(n))$ and the Hecke
algebra. The central result that we will use in the next section is the
statement that \AL\ generate a consistent solution of the YBE
on the reducible module $\oplus_{a=1}^{n-1} V_a$.

\chapter{A Quantum Group $S$-matrix}

An $S$-matrix constructed from the quantum group intertwiner discussed
in the last section will form the conjectured $S$-matrix describing
the solitons of the complex Toda field theory.
The deformation parameter $q$, in this context, is a coupling
constant. The ansatz, which was first written down in ref. [\Ref{tim}], is a
generalization of the soliton $S$-matrix for the sine-Gordon theory, where the
appropriate quantum group in that case is the deformation of $U(sl(2))$.
For the moment we consider the general problem of constructing
an $S$-matrix out of the $sl(n)$ quantum group intertwiner; the
connexion with the Toda theories will be addressed in \S4.

In general, the axioms of $S$-matrix theory do not strongly constrain
the form of the $S$-matrix. The situation in two dimensions for
integrable theories is exceptional [\Ref{zam}].
To start with, integrability implies
that there can be no particle creation or destruction, since the individual
momenta of the incoming particles must be conserved.
In addition an $N$-particle $S$-matrix can
be {\it factorized\/} as a product of
$\frac12N(N-1)$ 2-particle
$S$-matrices. Suppose that the particle states form a number of
degenerate multiplets labelled by a set of finite dimensional vector
spaces $\{V_a,\ a=1,2,\ldots,n-1\}$, with
masses $m_a$. Since the masses of the particles must be preserved on
scattering the only possible processes involve `flavour changing', {\it
i.e.\/} the $S$--matrix is determined by a set of maps
$$\sc^{a,b}(\theta):\ V_a\otimes V_b\longrightarrow V_b\otimes V_a,
\nfr{AP}
where $\theta=\theta_a-\theta_b$ is the rapidity difference of the
incoming particles. Consistency with factorization
implies that the two-body $S$-matrix must satisfy the
{\it Yang Baxter equation\/}
$$\eqalign{(I\ot\sc^{a,b}(\theta_1))&(\sc^{a,c}(\theta_1+\theta_2)\otimes
I)(I\otimes\sc^{b,c}(\theta_2))\cr =&(\sc^{b,c}(\theta_2)\otimes I)(I\otimes
\sc^{a,c}(\theta_1+\theta_2))(\sc^{a,b}(\theta_1)\otimes I),\cr}\nfr{AQ}
where $I$ is the identity operator on the appropriate vector space.
Further conditions come from the axioms of $S$--matrix theory and
integrability:

\noindent
\item{(i)} Unitarity\note{Unitarity in this context does not
imply that the underlying quantum field theory itself is unitary; that issue
depends on the signs of the residues of the $S$-matrix at particle
poles, an issue that we address in \S4.}.
$$\sc^{a,b}(\theta)\sc^{b,a}(-\theta)=I_b\otimes I_a,
\nfr{AR}
where $I_b\ot I_a$ is the identity in End$(V_b\ot V_a)$.

\noindent
\item{(ii)} Crossing symmetry.
$$\sc^{\ba,b}(\theta)=(I\otimes C_{a})\cdot\left(\sigma\cdot\sc^{b,a}
(i\pi-\theta)
\right)^{t_1}\cdot\sigma\cdot(C_{\bar a}\otimes I),\nfr{AS}
where $C_a:\ V_a\longrightarrow V_{\ba}$, ${\bar a}\in\{1,2,...,n-1\}$, is the
{\it charge conjugation operator\/}, and $C
_{\ba}C_a=I_a$ is the identity operator in $V_a$. As before,
$\sigma$ denotes the permutation, $\sigma(u\otimes v)= v\otimes u$, and
$t_1$ means `transpose' in the second space which is well defined because
$\sigma\cdot\sc^{b,a}(\theta)\in{\rm End}(V_b\otimes V_a)$.

\noindent
\item{(iii)} Analyticity. $\sc(\theta)$ is a meromorphic function of
$\theta$. The only singularities on the {\it physical strip\/}, $0
\leq{\rm Im}\,\theta\leq\pi$, are along
Re$\,\theta=0$ (since there can be no particle production for physical
values of the rapidity), and the simple poles correspond
to direct or cross-channel resonances.
If $\sc^{a,b}(\theta)$ has a simple pole at $\theta=iu_{ab}^c$ in the
direct channel we say that a particle of mass
$$m_c^2=m_a^2+m_b^2+2m_am_b\cos u_{ab}^c,
\nfr{AT}
is a bound state of $a$ and $b$. The new particle must itself be
included in the particle spectrum. If $ab\rightarrow c$ can occur then
so can $a{\bar c}\rightarrow{\bar b}$ and $b{\bar c}\rightarrow{\bar
a}$, where bar denotes charge conjugation. From \AT\ we deduce the
following identity
$$u_{ab}^c+u_{a{\bar c}}^{\bar b}+u_{b{\bar c}}^{\bar a}=2\pi.
\nfr{AU}

\noindent
\item{(iv)} The Bootstrap equations. The bootstrap equations give a
non--linear relation between $S$--matrix elements. If
$\sc^{a,b}(\theta)$ has a direct channel pole at $\theta=iu_{ab}^c$,
corresponding to a particle in $V_c=P_{ab}^c(V_a\ot V_b)$, where
$P_{ab}^c$ is a projection operator, then
$$\sc^{d,c}(\theta)=(I\ot\sc^{d,a}(\theta-i{\bar u}_{a{\bar c}}^{\bar b}
))(\sc^{d,b}(\theta+i{\bar u}_{b{\bar c}}^{\bar a})\ot I),
\nfr{AV}
restricted to $V_d\ot V_c\subset V_d\ot V_b\ot V_a$, and similarly
$$\sc^{c,d}(\theta)=(\sc^{b,d}(\theta-i{\bar u}_{b{\bar c}}^{\bar a})\ot
I))(I\ot \sc^{a,d}(\theta+i{\bar u}_{a{\bar c}}^{\bar b})),
\nfr{AW}
restricted to $V_c\ot V_d\subset V_b\ot V_a\ot V_d$. In the above ${\bar
u}_{ab}^c=\pi-u_{ab}^c$, {\it etc\/}.

The bootstrap constraints are very powerful because they must be
consistent with the integrability of the theory. What we mean by this is
that the spectrum of possible spins of the conserved charges is tightly
constrained and this in turn highly constrains the possible masses of
the physical states. There is a class of minimal solutions to the above
axioms for which each particle state is non-degenerate, {\it i.e.\/}
$V_a\simeq 1$, and the $S$-matrix has the minimum number of poles and
zeros needed to satisfy the axioms.
Each {\it minimal\/} solution is related to a simply-laced Lie
algebra. The minimal $S$-matrix (when multiplied by a function of the
coupling constant which introduces no additional poles onto the
physical strip) is then conjectured to be the $S$-matrix
of the particles of the affine Toda field theories for real coupling
constant discussed in the introduction.
The spins of the  conserved charges of a theory described by the
minimal $S$-matrix
are equal to the exponents of the finite Lie algebra $g$ modulo
its Coxeter number, and the number of particle states is equal to the
rank of $g$. So, for example, the particle spectrum of the $sl(n)$
theory consists of $n-1$ particles with masses
$$m_a=m_0\sin\left({\pi a\over n}\right),\ \ \ a=1,\ldots,n-1.
\nfr{AX}
Notice that these masses are, up to an overall scale factor, just the
classical masses of the fundamental Toda particles for real coupling
\classmass, but
also the classical soliton masses in the complex Toda theories \classsolmass.
The possible fusions are $ab\rightarrow(a+b)\ {\rm mod}\,n$, which
occur at the rapidity values $\theta=iu_{ab}$:
$$u_{ab}=\cases{{a+b\over n}\pi &$a+b<n$\cr \left(2-{a+b\over n
}\right)\pi\ \ &$a+b\geq n$.\cr}
\nfr{AY}
The charge conjugation operator maps $a\rightarrow{\bar a}=n-a$.
The explicit form for $\sc^{a,b}_{\rm min}(\theta)$, from ref.
[\Ref{arin}], is
$$\sc^{a,b}_{\rm min}(\theta)=(a+b)(a+b-2)^2(a+b-4)^2\cdots(|a-b|),
\nfr{minimalsm}
where the following notation has been borrowed from ref. [\Ref{todasm}]:
$$(x)={{\sin}\left({\theta\over2i}+{\pi x\over2n}\right)\over
{\sin}\left({
\theta\over2i}-{\pi x\over2n}\right)}.$$
The $S$-matrix element $S_{\rm
min}^{a,b}(\theta)$ has one direct channel pole at $\theta=i u_{ab}$
corresponding to the exchange of the particle $a+b\ {\rm mod}\,n$,
and a cross-channel pole at $\theta=i u_{a\bar b}$ corresponding to
the exchange of particle $a-b\ {\rm mod}\,n$.
Notice that since there is only one possible pole in
$\sc^{a,b}_{\rm min}(\theta)$ corresponding to a
bound state in the direct channel, we can unambiguously write
$u_{ab}\equiv u_{ab}^c$.

We now construct a series of factorizable $S$--matrices, with degenerate
particle states, based on the ansatz\note{We ignore
any ambiguities of the CDD type.}
$$\sc^{a,b}(\theta)=X^{a,b}(\theta)\rc^{a,b}(x(\theta)),
\nfr{AZ}
where $X^{a,b}(\theta)$ is a scalar function of $\theta$, and the
$\rc^{a,b}(x)$ are the solutions to the  YBE constructed in the last
section. It is convenient to split the scalar
prefactor $X^{a,b}(\theta)$ into
two pieces:
$$X^{a,b}(\theta)=S_{\rm min}^{a,b}(\theta)f^{a,b}(x(\theta)),
\efr
where $S_{\rm min}^{a,b}(\theta)$ is the minimal $S$-matrix written down in
\minimalsm.

The form of the ansatz identifies a particle of mass proportional to $m_a$
with the $a^{\rm th}$ fundamental representation of $U_q(sl(n))$. We now show
that the above ansatz is consistent with the axioms of $S$-matrix
theory. One should bear in mind that $\sc_{\rm min}^{a,b}(\theta)$, by
itself, satisfies all the axioms of $S$-matrix theory.

The YBE equation for $\sc(\theta)$ is satisfied because
$\rc(x)$ itself satisfies the YBE. However, we deduce that the spectral
parameter $x$ and the rapidity
$\theta$ must be related by $x=\exp(c\theta+d)$, for constants
$c$ and $d$.

Unitarity can be satisfied by an appropriate choice of
$f^{a,b}(x)$. Using \AO\ we deduce that $d=0$, {\it i.e.\/}
$x=\exp c\theta$ and
$$f^{a,b}(x)=\prod_{i,j=1}^{a,b}f(x(-q)^{-(a+b-2i-2j+2)/2}),\nfr{GGG}
where $f(x)$ is a scalar function satisfying
$$f(x)f(x^{-1})={1\over(q^{-1}x^{-1}-qx)(q^{-1}x-qx^{-1})}.\nfr{BA}

The $\rc^{a,b}(x)$ matrix must be consistent with the
bootstrap equations \AV\ and \AW.
In addition, the residue of $\sc^{a,b}(\theta)$
at the pole $\theta=iu_{ab}$, corresponding to the direct channel
bound state $c=a+b$, must be proportional to the projection operator
$s_{a+b}^-$. For example, $\sc^{1,1}(\theta)$ should be proportional to $s_2
^-$ when $\theta=iu_{11}=2\pi i/n$. By using equation \AE, we deduce
that
$$x=(-q)^{-n\theta/2\pi i}.
\nfr{BBB}
Now we know how $x$ is related to $\theta$ we can rewrite the bootstrap
equations in terms of the multiplicative variable $x$. It is easy to
show that the two bootstrap equations \AV\ and \AW\ then are identical to the
two fusion equations in \AL. In addition, equation \AN\ says that the
residue of $\sc^{a,b}(\theta)$ at the pole $\theta=iu_{ab}$ is
proportional to $s_{a+b}^-$, as required.

To exhibit crossing symmetry, we have to specify the
charge conjugation operator
$C_a:\ V_a\rightarrow V_{n-a}$.
In fact, $V_a$ is naturally dual to $V_{n-a}$ via the action of the
Hecke algebra analogue of the $\epsilon$-tensor, namely $s_n^-$.
This is because dim$(s_n^-(V_a\ot V_{n-a}))=1$,
where $V_a\ot V_{n-a}\subset V_1^{\ot\,n}$. From the action
$T({\bfmath e}_i\ot{\bfmath e}_j)={\bfmath e}_j\ot{\bfmath e}_i$, for
$i<j$ and equation \AD\ one finds
$$\eqalign{s_n^-({\bfmath e}_1\ot {\bfmath e}_2&\ot\ldots\ot{\bfmath e}_n)\cr
=&{q^{-n(n-1)/2}\over[n]!}\sum_{\{i_j\}\in P_n
}(-q)^{l(\{i_j\})}{\bfmath e}_{i_1}\ot{\bfmath
e}_{i_2}\ot\ldots\ot{\bfmath e}_{i_n},\cr}
\efr
where $P_n$ is the set of permutations of $\{1,2,\ldots,n\}$, and
$l(w)$ is the length of the permutation $w\in P_n$, with respect to
simple transpositions. If we define a set of dual vectors
$\{{\bfmath e}_i^\star,i=1,\ldots,n\}$, with ${\bfmath e}_i^\star\cdot
{\bfmath e}_j=\delta_{ij}$,
then the explicit expression for the charge conjugation operator is
$$C_a=\lambda_a\sum_{\{i_j\}\in P_n}(-q)^{l(\{i_j\})}
\left({\bfmath e}_{i_{a+1}}\ot\cdots\ot{\bfmath e}_{i_n
}\right)\left({\bfmath e}_{i_1}^\star\ot\cdots\ot{\bfmath
e}_{i_a}^\star\right).
\nfr{HHH}
In the above, $\lambda_a$ is a normalization constant determined by $C
_{\bar a}C_a=I_a$. This gives
$$\lambda_{\bar a}\lambda_a={(-)^{a{\bar a}}q^{-a(a-1)/2-{\bar a}({\bar
a}-1)/2-a{\bar a}}\over[a]![{\bar a}]!}.$$
For the case $n=2$, corresponding to the sine-Gordon
theory, the action of the charge conjugation operator is unconventional.
In the usual formulation of the soliton $S$-matrix in
sine-Gordon theory, the ansatz of \AZ\ needs to be conjugated with
momentum dependent factors in order to ensure crossing symmetry
[\Ref{sgle}]. However, in the present formulation this is not required
because the charge conjugation operator acts in an unconventional way:
$$
C{\bfmath e}_1=(-q)^{\frac12}{\bfmath e}_2,\qquad C{\bfmath e}_2=
(-q)^{-\frac12}{\bfmath e}_1,
\nfr{char}
to compare with the usual action $C{\bfmath e}_1={\bfmath e}_2$ and
$C{\bfmath e}_2={\bfmath e}_1$. The unconventional action \char\
makes the introduction of momentum depend factors in the definition of
the $S$-matrix unnecessary, but is obviously equivalent to the usual
formulation by a redefinition of the in and out states.

To determine the constraint implied by crossing symmetry, it is
sufficient to consider the relationship between $S^{1,1}(\theta)$ and
$S^{n-1,1}(\theta)$. It is possible to show directly that the
$S$-matrix elements are related by crossing symmetry if the following
equation is satisfied by the function $f(x)$:
$$\prod_{i=1}^{n-1}f(x^{-1}(-q)^{-i})
\prod_{i=1}^{n-2}(x^{-1}(-q)^{-i-1}-x(-q)^{i+1})=f(x).
\nfr{BB}

\def\mw{{in\lambda\theta\over2\pi}}

To complete the construction of a consistent $S$-matrix, we must find
a function $f(x)$, which satisfies \BA\ and \BB.
Introducing the notation $q=-\exp-i\pi\lambda$,
the explicit expression for $f(x)$, derived in appendix B, is
$$\eqalign{&f(x)={\Gamma\left(\mw+\lambda\right)\Gamma\left(1-\mw-\lambda
\right)\over2\pi i}
\prod_{j=1}^\infty{\Gamma\left(1+\mw+(j-1)n\lambda\right)\over\Gamma
\left(1-\mw+(j-1
)n\lambda\right)}\cr &
\times{\Gamma\left(\mw+jn\lambda\right)\Gamma
\left(-\mw+((j-1)n+1
)\lambda\right)\Gamma\left(1-\mw+(jn-1)\lambda\right)\over
\Gamma\left(-\mw+jn\lambda\right)\Gamma\left
(\mw+((j-1)n+1)\lambda\right)\Gamma\left(1+\mw+(jn-1)\lambda\right)}.\cr}
\nfr{TTT}

To summarize: the fusion procedure for the $R$-matrix provides a solution
to the bootstrap equations if we identify the particle of mass
proportional to $m_a$
with the fundamental representation $\varrho_a$ of $U_q(sl(n))$.

\chapter{The Soliton $S$-Matrix}

The quantum group $S$-matrix in equation \AZ\
represents our ansatz for the soliton-soliton $S$-matrix.
We now consider some of the implications of this proposal.
First of all, the fact that the poles of the minimal part of the
$S$-matrix encode the fusing between the solitons implies the quantum
soliton masses are proportional to \AX. This means that, up to an
overall multiplicative renormalization, the quantum soliton masses
$\hat M_a$ are proportional to the classical soliton masses $M_a$ in
\classsolmass, or the masses of the fundamental Toda particles $m_a$.

The topological charge operator is proportional to
the Cartan subalgebra generator $h$ of $U_q(sl(n))$, which has the
following action on $V_1$:
$$h{\bfmath e}_i=e_i\,{\bfmath e}_i,
\efr
where $e_i$ is one of the weights of the $n$ dimensional
representation.
It is immediately apparent that
$$
[h\ot I+I\ot h,\rc(x)]=0,
\efr
and so the proposed $S$-matrix conserves topological charge. In the
quantum theory there is a soliton state for every weight of a
fundamental representation. In the classical scattering theory, not
all the weights of the fundamental representations
are obtained (except for the $n$ and $\bar n$ representations).

The fact that the {\it fusing rules\/} of the solitons \AY\ are
exactly the same as the those of the fundamental Toda particles in the
real coupling constant theories, implies that the spectrum of
conserved charges is the same.
All these conserved charges are scalars with
respect to the quantum group. However, there exist symmetries which
are non-trivial with respect to the quantum group. These
`residual quantum symmetries' generalize the situation for the
sine-Gordon theory [\Ref{sgle},\Ref{sgpert}].
To see this in more detail we have to
introduce the generators of $U_q(sl(n))$,
$\{e_i,f_i,h_i,i=1,\ldots,n-1\}$\note{Here, $h_i=\alpha_i\cdot h$, where
$h$ is the Cartan subalgebra generator introduced above.}. (The
$e_i$'s should not be confused with the weights of the $n$ dimensional
representation.)
The algebra of the generators may be found in ref. [\Ref{jimbo}].
Recall that $\rc(x)$
is invariant under the action of the quantum group whose action is
defined by the comultiplication $\Delta^{(2)}$:
$$\eqalign{\Delta^{(2)}(e_i)&=q^{h_i/2}\ot e_i+e_i\ot q^{-h_i/2}\cr
\Delta^{(2)}(f_i)&=q^{h_i/2}\ot f_i+f_i\ot q^{-h_i/2}\cr
\Delta^{(2)}(h_i)&=h_i\ot I+I\ot h_i.\cr}
\efr
So crudely speaking the $S$-matrix is quantum group invariant.
However, the $\rc(x)$ matrix also enjoys a momentum
dependent symmetry. To see this we note that the $n$ dimensional
representation of
$U_q(sl(n))$ is identical to that of $sl(n)$. Let $(e_0,f_0)$ be the
raising and lowering operators associated to the highest root. $\rc(x)$
satisfies
$$\eqalign{\rc(x_1/x_2)(x_1^2e_0\ot q^{-h_0/2}+&x_2^2q^{h_0/2}\ot e_0
)\cr =&(x_2^2e_0\ot q^{-h_0/2}+x_1^2q^{h_0/2}\ot e_0)\rc(x_1/x_2),\cr}
\nfr{fracgen}
where $h_0=-\sum_{i=1}^{n-1}h_i$. A similar equation holds with $e_0$
replaced by $f_0$ and $x_1$ and $x_2$ interchanged. We
interpret the above as a momentum dependent
symmetry of the $S$-matrix. The spin of the generator of
such a symmetry follows from the relation between $\theta$ and  $x$ in
equation \BBB. With $q=\exp-\pi i\lambda$,
the spin  of the generator is $n\lambda$.

Since we have the full $S$-matrix of solitons we can discover whether there
are any additional states in the theory, which will
manifest themselves as simple
poles on the physical strip. These new states must
then be added into the list of states of the
theory. The bootstrap equations can then be used to extract the
$S$-matrix elements of these additional states. For instance, in the
case of $sl(2)$, the sine-Gordon theory, one finds a set of poles,
whose position depends on the coupling constant, corresponding to the
exchange of scalar states. These are the `breathers' or `doublets'.
The masses of the breathers, as written down in \massbr, follow from
the positions of the poles.
So we must search for simple poles on the
physical strip, over and above the simple poles already interpreted as
being due to the exchange of solitons.

To begin with, we consider just
the scattering of elementary solitons. There are three
possible classes of process to consider. (i) The transmission
process\note{The notation here is short for
$(e_i,e_k)\rightarrow(e_k,e_i)$.}, $ik\rightarrow ki$,
for $i\neq k$. (ii) The identical particle
process, $ii\rightarrow ii$. (iii) The reflection process,
$ik\rightarrow ik$. These processes are illustrated in figure 3.

\midinsert
\centerline{Figure 3. Transmission, reflection and identical particle
processes.}
\endinsert

The $S$-matrix elements for the transmission and identical particle
processes, after some re-arranging of
the arguments of the gamma functions using the product representation
of the gamma function, are
\def\aa{{i\theta\over2\pi}}
$$\eqalign{
S_{ik\rightarrow ki}(\theta)=\prod_{j=1}^\infty&{\Gamma\left(\aa+1+{j\over
n\lambda}\right)\Gamma\left(\aa+1+{j-1\over n\lambda}\right)\over
\Gamma\left(-\aa+{j\over n\lambda}\right)\Gamma\left(-\aa+{j-1\over n
\lambda}\right)}\cr
&\quad\quad\times{\Gamma\left(-\aa+{1\over n}+{j\over n\lambda}\right)
\Gamma\left(-\aa-{1\over n}+{j-1\over n\lambda}\right)\over
\Gamma\left(\aa+1+{1\over n}+{j\over n\lambda}\right)\Gamma\left(\aa+1
-{1\over n}+{j-1\over n\lambda}\right)},\cr}
\nfr{ijji}
and
$$
S_{ii\rightarrow
ii}(\theta)={\sin\left(\pi\lambda-{in\lambda\theta\over2}\right)\over
\sin\left({in\lambda\theta\over2}\right)}S_{ik\rightarrow ki}(\theta).
\nfr{iiii}
Consider the $S$-matrix
element $S_{ik\rightarrow ki}(\theta)$. It has simple poles on the
physical strip ($0\leq{\rm Im}(\theta)\leq\pi$) at
$$
\theta=-{2\pi i\over n\lambda}p+{2\pi i\over n},\qquad p=0,1,\ldots,
[\lambda].
\nfr{pijji}
In the above, the
notation $[\lambda]$ means the largest integer less than $\lambda$.
The simple pole at $\theta=2\pi i/n$, corresponds to the
exchange of a soliton with topological charge $e_i+e_k$ and mass $\hat
M_2$ in the direct channel.
The residues of the poles in \pijji\ are proportional to $s_2^-$, and
so they are
interpreted as being due to the exchange of excited solitons, or
`breathing solitons', in the direct channel with topological charge
$e_i+e_k$. The soliton corresponding to the simple pole at
$\theta=2\pi i/n$, is just the ground
state of this system. The masses of the excited solitons are
$$
\hat M_2(p)=2\hat M_1\cos\left({\pi\over n}\left(1-{p\over\lambda}
\right)\right),\qquad p=0,1,\ldots,\left[\lambda\right],
\nfr{excitsolmass}
where $\hat M_2=\hat M_2(0)$. For the sine-Gordon theory the poles
\pijji\ and the masses have a different interpretation. In this case
the states being exchanged are the breathers; their topological charge
is zero and the masses in \excitsolmass\ are equal to the breather
masses \massbr\ (with the identification $\gamma=4\pi/\lambda$).

Similarly the element $S_{ii\rightarrow ii}(\theta)$ has simple poles
on the physical strip at
$$
\theta={2\pi i\over n\lambda}p,\qquad
p=1,2,\ldots,\left[\frac12n\lambda\right].
\nfr{piiii}
These poles are at $x=e^{-\pi ip}$ for which
$$
\rc(x=e^{-i\pi p})=(-)^p(q-q^{-1})I\otimes I,
\efr
and so it is natural to interpret these poles as being due to the
exchange of scalar states ({\it i.e\/}. with zero topological charge) in the
cross channel of $ii\rightarrow ii$: these scalar states are the
analogues of the sine-Gordon breathers. The masses of the states are
$$
\hat m_1(p)=2\hat M_1\sin\left({\pi p\over n\lambda}\right),\qquad
p=1,2,\ldots,\left[\frac12n\lambda\right],
\nfr{massgenbr}
which generalizes the sine-Gordon result \massbr.

We now consider the other $S$-matrix elements $S^{a,b}(\theta)$, where
we take $a+b\leq n$ and $a\geq b$, without loss of generality.
Rather than give a full discussion, we only present the
results; a complete analysis will appear elsewhere.
There are two types of simple pole, those whose positions do not depend on the
coupling constant and those whose positions do. In the first set are
$$
\theta={\pi i\over n}(a+b),\quad{\pi i\over n}(a-b),
\nfr{fixpol}
which correspond to the soliton $a+b$, in the direct channel, and
$\bar a+b$, in the crossed channel. In the second set are
$$\theta=-{2\pi i\over n\lambda}p+{\pi i\over n}(a+b-2j+2),\qquad
p=1,2,\ldots,\left[\frac12\lambda(a+b-2j+2)\right],
\nfr{dirpol}
with $j=1,2,\ldots,b$, and
$$\theta={2\pi i\over n\lambda}p+{\pi i\over n}(a+b-2j),\qquad
p=1,2,\ldots,\left[\frac12\lambda(2j-a-b+n)\right],
\nfr{cropol}
also with $j=1,2,\ldots,b$. The poles in \dirpol\ are
direct channel poles and those in \cropol\ are crossed channel poles.
The poles correspond to solitons transforming in
non-fundamental representations, excited solitons in both
fundamental and non-fundamental representations, and scalar states.
The excited solitons corresponding to the fundamental representation
$\varrho_{a+b}$ have masses given by the square root of
$$
\hat M_a^2+\hat M_b^2+2\hat M_a\hat M_b\cos\left({\pi\over n}
\left(a+b-{2p\over\lambda}\right)
\right),\qquad p=1,2,\ldots,\left[\frac12\lambda(a+b)\right],
\nfr{exgensol}
which generalizes \excitsolmass. The scalar states correspond to
simple poles in the
direct channel of $S^{\bar a,a}(\theta)$, and have masses
$$
\eqalign{
\hat m_a(p,j)=&2\hat M_a\sin\left({\pi \over n}\left({p\over\lambda}
+j-1\right)\right),\cr &\qquad j=1,2,\ldots,a,\quad
p=1,2,\ldots,\left[\frac12\lambda(n-2j+2)\right],\cr}
\nfr{genscal}
which generalizes \massgenbr.

Another issue concerns the question as to whether the $S$-matrix
describes a unitary quantum field theory.
As we have already pointed out, this is a
separate issue from whether the $S$-matrix is unitary as a matrix. A
discussion of this point may be found in ref. [\Ref{cardy}] which
discusses the non-unitary field theory describing the Lee-Yang edge
singularity. Consider the $S$-matrix element describing the scattering
of two equal mass particles. If $S$ has a pole at $\theta=iu$
corresponding to the exchange of particle in the direct channel, then
for a unitary theory one has
$$
S(\theta)\sim{i\rho^2\over\theta-iu},
\efr
for some $\rho\in{\Bbb R}$. For a non-unitary theory the residue
might have a different sign. For example, consider the excited soliton
poles of equation \pijji\  for the process
$ik\rightarrow ki$. The explicit expression for the $S$-matrix element
for this process is equation \ijji. The pole corresponding to the
$p^{\rm th}$ excited soliton comes from the factor
$$
\Gamma\left(-{i\theta\over2\pi}-{1\over n}+{p\over n\lambda}\right).
$$
The only `dangerous' terms as regards the sign of the residue are
$$
\Gamma\left(-{p\over n\lambda}\right)\Gamma\left(-{p-1\over n\lambda}\right)
\cdots\Gamma\left(-{1\over n\lambda}\right),
$$
which contributes an overall sign of $(-)^p$ to the residue.
So the first excited soliton corresponds to a non-unitary
coupling, the second a unitary coupling {\it etc\/}. The message of this
result is that the underlying quantum field theory is, in general,
non-unitary. This was only to be expected, considering the complex form for
the Hamiltonian \ham. However, as we have seen for the elementary
solitons, there exists a regime
for which all the non-unitary states decouple. In the next section we
make some comments about this for the full $S$-matrix.

The $S$-matrix reduces to a very simple expression when $\lambda=1$.
In this case $q=1$ and so the Hecke algebra reduces to the symmetric
group and the $S$-matrix becomes trivial in the space of topological
charges. One can easily verify from the explicit expressions that
$$
\left.S^{a,b}(\theta)\right\vert_{\lambda=1}=S_{\rm
min}^{a,b}(\theta).
\efr

Finally let us
consider the case for $n=2$, where the $S$-matrix is the soliton
$S$-matrix of the sine-Gordon theory. In this case $\sc_{\rm
min}(\theta)$ has no poles on the physical strip; it is a CDD type
ambiguity and so for this
case alone we drop this part of the $S$-matrix without
affecting any properties of the ansatz. The resulting $S$-matrix is
exactly that of ref. [\Ref{zam}], up to the unconventional action of
the charge conjugation operator \char.

\chapter{The Semi-Classical Limit}

In this section we will verify that some particular elements of the
conjectured $S$-matrix are consistent with the classical scattering
theory. The idea is to use the relation between the semi-classical
limit of the $S$-matrix (the limit as $\hbar\rightarrow0$)\note{In
this section we shall re-introduce $\hbar$ into our formulas.}
 and the
time-delays of the classical scattering theory. We will only consider
the scattering of the elementary solitons (the $n$-dimensional
representation) for simplicity.

As we discussed in \S3 there are only
three possible classes of process involving elementary solitons. The
transmission, identical particle and reflection processes.
It turns out that
only the semi-classical limit of the first two processes can be
connected in a simple way with the classical scattering theory. A
discussion of the reflection amplitude requires an analysis of a
complex time trajectory in the classical theory [\Ref{reflect}], which we will
postpone for a future publication.

For the transmission and identical particle processes, there
is a very simple relation between the leading term of the
semi-classical limit of the $S$-matrix element and the corresponding
time-delay of the classical theory. We simply quote the result which
follows from the WKB analysis of ref. [\Ref{semiclass}]. If $E$ is
the energy in the channel in question and $\Delta t(E)$ is the
classical time-delay, defining
$$
\delta(E)=\frac12n_B\pi+\frac12\int_{E_{\rm th}}^EdE'\,\Delta t(E'),
\nfr{wkbform}
where the number of bound states, or resonances,
in the channel is the largest integer
less than $n_B$ (which is denoted $[n_B]$), and $E_{\rm th}$ is the
threshold energy where the resonances are just unbound, then the
leading behaviour of the $S$-matrix is
$$
S(\theta)=\exp\,{2i\over\hbar}\left(\delta(\theta)+O(\hbar)\right).
\nfr{smatexpOB}
Let us apply \wkbform\ to the scattering of elementary solitons. In
the centre-of-mass the total energy is $E=2M_1{\rm cosh}(\theta/2)$,
where $\theta$ is the relative rapidity. The threshold energy is
$2M_1$. The classical time-delay is from \S1
$$
\Delta t(\theta)=-{2n\gamma^{(12)}(\theta)\over M_1\beta^2{\rm
sinh}(\theta/2)}.
\efr
Writing the integral \wkbform\ in terms of the rapidity we have
$$\delta(\theta)=\frac12n_B\pi-{n\over\beta^2}\int_0^\theta d\theta'\,
\gamma^{(12)}(\theta').
\nfr{phase}
In \phase, $\gamma^{(12)}(\theta)$ is the `interaction
function' \intconst\ for the classical soliton scattering.
For two elementary solitons we have
$$\gamma^{(12)}(\theta)=\log\left({\sin^2\left({\theta\over2i}\right)
\over\sin\left({\theta\over2i}+{\pi\over n}\right)\sin\left({\theta
\over2i}-{\pi\over n}\right)}\right).
\nfr{intel}
The two relevant $S$-matrix elements are written down in \ijji\ and
\iiii. In order to implement the semi-classical limit we have to know how
the coupling constant $\lambda$ (or $q$) depends on $\hbar$. We have
already seen that the limit $q\rightarrow1$ can be thought of as the
weak coupling limit for the solitons, since the $S$-matrix becomes
trivial, however, this cannot correspond to the limit
$\beta\rightarrow0$ because the soliton masses are proportional to
$\beta^{-2}$. This situation is familiar from the sine-Gordon theory,
the point being that the solitons are weakly coupled when the coupling
constant $\beta$ is large. Mirroring the situation of the sine-Gordon
theory, we shall find that a behaviour of the form
$$
\lambda={1\over\hbar\beta^2}(\lambda_0+O(\hbar\beta^2)),
\nfr{lamhbar}
will be necessary. So as $\hbar\rightarrow0$
$\lambda\rightarrow\infty$. In this limit the product of the gamma
functions in \ijji\ may be approximated by an integral in the
exponent, so the leading order behaviour is
$$\eqalign{
S_{ik\rightarrow ki}(\theta)\rightarrow&\exp\left\{{n\lambda_0
\over\hbar\beta^2}\int_0^\infty
dx\,\log\left({\Gamma\left(\aa+1+x\right)\Gamma\left(\aa+1+x\right)\over
\Gamma\left(-\aa+x\right)\Gamma\left(-\aa+x\right)}\right.\right.\cr
&\qquad\left.\left.{\Gamma\left(-\aa+{1\over n}+x\right)\Gamma\left(
-\aa-{1\over n}+x\right)\over\Gamma\left(\aa+1+{1\over n}+x\right)
\Gamma\left(\aa+1-{1\over n}+x\right)}\right)\right\}.\cr}
\efr
But this is equal to
$$\eqalign{
&\exp\left\{{n\lambda_0\over\hbar\beta^2}\int_0^{1\over n}dx\,\log\left(
{\Gamma\left(\aa+1+x\right)\Gamma\left(-\aa-x\right)\over
\Gamma\left(\aa+1-x\right)\Gamma\left(-\aa+x\right)}\right)\right\}\cr
=&\exp\left\{{n\lambda_0\over\hbar\beta^2}\int_0^{1\over n}dx\,\log\left(
{\sin\left(-\pi x-{i\theta\over2}\right)\over\sin\left(\pi x-{i\theta
\over2}\right)}\right)\right\}\cr
=&\exp\left\{i\pi{\lambda_0\over\hbar\beta^2}+{in\lambda_0\over2\pi\hbar
\beta^2}
\int_0^\theta d\theta'\,\log\left({\sin\left({\theta'\over2i}+{\pi\over
n}\right)\sin\left({\theta'\over2i}-{\pi\over n}\right)\over
\sin^2\left({\theta'\over2i}\right)}\right)\right\}.\cr}
\efr
{}From, this we can extract the phase shift
$$
\delta_{ik\rightarrow
ki}(\theta)={\lambda_0\pi\over2\beta^2}+{n\lambda_0\over4\pi\beta^2}
\int_0^\theta d\theta'\,\log\left({\sin\left({\theta'\over2i}+{\pi\over
n}\right)\sin\left({\theta'\over2i}-{\pi\over n}\right)\over
\sin^2\left({\theta'\over2i}\right)}\right).
\nfr{deltamis}
Now compare the above with \phase\ and the expression for
$\gamma^{(12)}(\theta)$ in \intel. The two expressions agree if
$$
\lambda_0={4\pi},
\nfr{lambdazero}
and $n_B$, the parameter which relates to the number of bound states
in the channel $ik\rightarrow ki$, is equal to $\lambda$ (at this
order in $\beta^2$). This is consistent with the number of bound
states found in the direct channel of this process ($\sim\lambda$)
found in \S4.

In a similar way we can repeat the analysis for the process
$ii\rightarrow ii$. However, it is more straightforward to take the
semi-classical limit of \iiii
$$
S_{ii\rightarrow ii}(\theta)=S_{ik\rightarrow ki}(\theta)\exp\left(
-i\pi{\lambda_0\over\hbar\beta^2}\right),
\efr
from which we deduce
$$
\delta_{ii\rightarrow ii}(\theta)={n\lambda_0\over4\pi\beta^2}
\int_0^\theta d\theta'\,\log\left({\sin\left({\theta'\over2i}+{\pi\over
n}\right)\sin\left({\theta'\over2i}-{\pi\over n}\right)\over
\sin^2\left({\theta'\over2i}\right)}\right).
\nfr{delta}
Again this is the correct semi-classical limit if $n_B$, in this case,
is zero. This is consistent with the findings of \S4, where we found no
bound states in the direct channel of the process $ii\rightarrow ii$
(however, there are bound states in the crossed channel).

So we have shown, at least for some of the processes,
that the conjectured $S$-matrix with the following functional
form for $q$
$$
q=\exp-{1\over\hbar\beta^2}\left(4\pi^2i+O(\hbar\beta^2)\right),
\nfr{functformq}
does indeed describe the quantization of the classical scattering
theory. In order to find the higher order terms in \functformq\ it
would be necessary to go beyond the WKB approximation. Such an
analysis has been done for the sine-Gordon theory in refs.
[\Ref{wkb},\Ref{jae}] yielding the result
$$
q=\exp\left(-{4\pi^2i\over\hbar\beta^2}+\pi i\right),
\efr
or equivalently $\gamma=4\pi/\lambda$, where $\gamma$ is defined in
\defgam.

\chapter{Discussion}

We have constructed the classical scattering theory of the solitons of
the complex affine $sl(n)$ Toda equations. The solitons have
topological charges which are weights of the fundamental
representations of the Lie algebra $sl(n)$, and masses whose ratios
are equal to those of the conventional Toda particles.
Using the sine-Gordon
theory as a paradigm, we proposed a form for the $S$-matrix of the
solitons by generalizing the $sl(2)$ quantum group to the $sl(n)$
quantum group. The quantum group acts in space of topological charge
of the quantum soliton theory and the
resulting $S$-matrix commutes with its action,
as well as with a momentum dependent operator
leading to non-trivial conserved quantities over and above the ones
associated to a real Toda theory. The ansatz related the
coupling constant of the Toda theory to the deformation parameter of
the quantum group \functformq. The quantum masses of the solitons are
proportional to their classical masses, up to an overall
renormalization. This is consistent with the lowest order quantum
corrections to the soliton masses, which can be computed in the
following way. One looks at the linearized equation around the soliton
solutions. The equation is a multi-component Schr\"odinger equation.
Surprisingly, given the fact that the potential is complex, the
frequencies of the modes are real, and hence the solitons are stable
to small perturbations. Furthermore, the zero-point energies of the
modes may be summed to give the lowest order correction to the soliton
masses. Details of this calculation will be presented elsewhere.

The spectrum of the $sl(n)$ theories, for $n>2$, is a good deal more
complicated than the sine-Gordon theory. In addition to the scalar
breather states, there are also excited soliton states. Ideally, one
would like to construct the $S$-matrix elements of these new states by
applying the bootstrap equations. For the sine-Gordon theory this
procedure terminates, no new states are then produced, and the
complete spectrum just consists of the soliton anti-soliton and
breathers. The problem of finding the complete spectrum for the
general soliton theories looks rather formidable.
The $S$-matrix describes a non-unitary theory, as expected from the
complex form for the Hamiltonian. For the
sine-Gordon theory the ground state of the breather is identified with
the `elementary particle' of the theory. We now show that such an
identification can be made for the $sl(n)$ theories.
If we expand the equations of motion \eqnmotion\ in powers of $\phi$
then in the linear approximation we expect to see modes corresponding
the `classical' masses of \classmass, as in the real Toda theories. On
quantizing we might expect these modes to appear as quanta with mass
$m_a+O(\beta^2)$,
where the corrections arise from loops. Indeed, we did find scalar
particles with masses given by \genscal. For each $a=1,2,\ldots,n-1$
there is a discrete spectrum of particles. The ground states of these
spectra have mass
$$\hat m_a(1,1)=2\hat M_a\sin\left({\pi\over n\lambda}\right),
\efr
In the limit of weak
coupling, or $\hbar\rightarrow0$, using the expression for the
classical soliton masses and the expression for $\lambda$ we deduce
$$
\hat m_a(1,1)=2m\sin\left({\pi a\over n}\right)+O(\beta^2),
\efr
which are the masses of the elementary particles in the weak
coupling limit.

{}From the point of view of the $S$-matrix $S^{a,b}(\theta)$, there are
two types of particle appearing as bound states, depending on whether
the position of the associated pole depends on the coupling constant \fixpol,
or not, \dirpol\ and \cropol. In the former set, there are only the original
solitons themselves, associated to the fundamental representations.
The latter set contains the excited solitons, the solitons corresponding to
non-fundamental representations and the scalar particles.
There is a region for the coupling constant, namely
$$\lambda<{2\over n},
\nfr{regio}
for which all the poles corresponding to the latter states are no
longer on the physical strip. Remarkably the resulting $S$-matrix now
describes a unitary quantum field theory, since all the non-unitary
couplings are associated with the second set of states. Hence,
for \regio\ the soliton $S$-matrix is complete and describes a
unitary theory. Full details of this will be presented elsewhere.

For the sine-Gordon theory, $\lambda<1$ is
equivalent to $\gamma>4\pi$, or $\beta^2>2\pi$, which is the repulsive
regime, for which the $S$-matrix, as presented, is no longer valid. It
would clearly be of interest to discover whether the general Toda
theories have an analogue of the repulsive regime.

Returning to the $S$-matrix constructed in \S2, it is easy to see that the
axioms of $S$-matrix theory are satisfied
algebraically (from the point of view of the Hecke algebra). This
means that they will hold for any representation of the
Hecke algebra. In particular, it means that we can formulate the theory
in the {\it Interaction Round a Face\/} (IRF), or {\it
Solid-On-Solid\/} (SOS), picture. From a soliton point of view this corresponds
to labelling
the processes by specifying the vacua between the solitons, rather
than the topological charges of the solitons. Obviously this is
completely equivalent to the `vertex' point of view adopted in \S2. However,
this equivalence is only true for generic values of $q$ assumed in
\S2. When $q$ is a root of unity, say $\lambda=1/p$, where $p\in{\Bbb Z}>n$,
the `vertex' description is no longer
appropriate since some of the elements of the $R$-matrix become
singular. In the `face'
picture, however, there is a way of restricting the allowed variables
to a finite set for which the $R$-matrix is well-defined and the Hecke
algebra relations are still satisfied. In terms of the solitons this
would correspond to restricting the allowed set of vacua to some
finite set (recall that the allowed set of vacua is isomorphic to the
weight lattice of $sl(n)$). For these restricted models the spin of
the symmetry generator in \fracgen\ is equal to $n/p$, hence these
theories have fractional symmetries generalizing the situation for the
restricted sine-Gordon theories in refs. [\Ref{sgpert}]. Notice that
the restricted $S$-matrices would lie in the regime \regio, so they
would describe unitary quantum field theories.

In fact there are more general representations of the Hecke algebra
associated to certain graphs [\Ref{Vr}] to which one could also
associate a factorizable soliton $S$-matrix. In addition, one could
also consider quantum groups related to other Lie algebras.

\acknowledgements
I would like to thank Jonathan Evans for some helpful conversations
regarding the reality of the Toda equations.
I also acknowledge the support of Merton College, Oxford, through a Junior
Research Fellowship.

\appendix{Appendix A}

In this appendix we prove the result that the $\rc^{a,b}(x)$ matrices
defined in equation \AL\ satisfy the YBE. As we have already
stated in the text the fusion procedure will guarantee that the YBE if
\AN\ is satisfied. For reasons of space, the proof that we present
is not self-contained, since we shall assume two
lemmas which may be found in the second reference of [\Ref{jimbo}].

In the following we use the notation $\rc_a(x)=I\ot\cdots \ot I
\ot\rc(x)\ot I\ot\cdots\ot I$ to denote the basic $\rc(x)$ acting between the
$a^{\rm th}$ and $(a+1)^{\rm th}$ space in the tensor product $V_1\ot\cdots\ot
V_1$, and $s^-_{a,b}$ to denote the Hecke algebra analogue of the full
anti-symmetrizer $s_{b-a+1}^-$
acting between the $a^{\rm th}$ space and the $b^{\rm th}$ space,
inclusive, in the tensor product $V_1\ot\cdots\ot V_1$.

\noindent
$\underline{\rm Lemma\ 1.}$ Equation (9) of the second reference of
[\Ref{jimbo}].
$$\eqalign{\rc_{a}(x)\rc_{a-1}(x(-q)^{-1})\cdots\rc_{1}&
(x(-q)^{-a+1})s^-_{2,a+1}\cr
=&s^-_{1,a}\rc_{a}(x(-q)^{-a+1})\rc_{a-1}(x(-q)^{-a+2})\cdots\rc_{
1}(x),\cr}$$
and similarly
$$\eqalign{\rc_{1}(x)\rc_{2}(x(-q)^{-1})\cdots\rc_{a}&
(x(-q)^{-a+1})s^-_{1,a}\cr
=&s^-_{2,a+1}\rc_{1}(x(-q)^{-a+1})\rc_{2}(x(-q)^{-a+2})\cdots\rc_{a
}(x).\cr}$$
$\underline{\rm Lemma\ 2.}$
$$\eqalign{s^-_{1,a+1}&={1\over (-q)^{-a-1}-(-q)^{a+1}}s^-_{1,a}\rc_{a}
((-q)^{-a})s^-_{1,a}\cr &={1\over
(-q)^{-a-1}-(-q)^{a+1}}s^-_{2,a+1}\rc_{1}((-q)^
{-a})s^-_{2,a+1}.\cr}$$
By applying Lemma 2 $b$ times one can easily show
$$s_{1,a+1}^-\propto s^-_{1,a}\rc_a((-q)^{-a})\rc_{a-1}((-q)^{-a+1})\cdots
\rc_{a-b+1}((-q)^{-a+b-1})s^-_{1,a-b+1},
\nfr{onef}
and
$$s_{1,a+1}^-\propto s_{2,a+1}^-\rc_1((-q)^{-a})\rc_2((-q)^{-a+1})\cdots
\rc_b((-q)^{-a+b-1})s^-_{b+1,a+1}.
\nfr{twof}
We now turn to the main proof.
$\rc^{a,b}(x)$ is defined recursively in terms of $\rc(x)$ by using the fusion
equations \AL. What we need to show is that the
$\rc^{a,b}(x)$ so defined satisfies \AN:
$$\rc^{a,b}\left((-q)^{-(a+b)/2}\right)\propto s^-_{a+b}.
\nfr{proj}
We proceed by induction.
Firstly, equation \AE\ implies that \proj\ is satisfied for $a=b=1$:
$$\rc^{1,1}((-q)^{-1})\equiv\rc((-q)^{-1})\propto s_2^-.$$
Now consider
$$\eqalign{\rc^{a+1,b}&\left((-q)^{-(a+b+1)/2}\right)\cr
=&(\rc^{a,b}((-q)^{-(a+b)/2})\ot I)
(I\ot\rc^{1,b}((-q)^{-a-(b+1)/2}))(s^-_{a+1}\ot s_b^-),\cr}
\nfr{next}
where we have introduced the projection operator $(s^-_{a+1
}\ot s^-_{b})$ to implement the restriction $V_{a+1}\ot V_b\subset
V_1^{\ot(a+b+1)}$, explicitly. Assuming that \proj\ is true for
$\rc^{a,b}(x)$ we can write the right--hand side of \next\ as
$$(s^-_{a+b}\ot I)(I\ot\rc^{1,b}((-q)^{-a-(b+1)/2}))(s^-_{a+1}\ot
s^-_b).
\nfr{ohhh}
By applying \AL\ $b-1$ times we can express $\rc^{1,b}(x)$ in terms of
the basic $\rc(x)$
$$\rc^{1,b}(x)=\rc_{b}(x(-q)^{(b-1)/2})\rc_{b-1}(x(-q)^{(b-3)/2})
\cdots\rc_{1}(x(-q)^{-(b-1)/2})s_{2,b+1}^-.
\nfr{helpp}
Using Lemma 1 we can rewrite the right--hand side of \helpp\ as
$$s_{1,b}^-\rc_{b}(x(-q)^{-(b-1)/2})\rc_{b-1}(x(-q)^{-(b-3)/2})
\cdots\rc_{1}(x(-q)^{(b-1)/2}).$$
Substituting this into \ohhh\ we find
$$s^-_{1,a+b}\rc_{b+a}((-q)^{-a-b})\rc_{b+a-1}((-q)^{-a-b+1})
\cdots\rc_{a+1}((-q)^{-a-1})s_{1,a+1}^-,$$
where we have using the fact that $s_{a,b}^-s^-_{c,d}=s^-_{a,b}$ for
$a\leq c<d\leq b$. But by \onef, the corollary of Lemma 2, this is
proportional to $s_{a+b+1}^-$, as required.

To complete the proof we must consider
$$\eqalign{\rc^{a,b+1}&\left((-q)^{-(a+b+1)/2}\right)\cr
=&(I\ot\rc^{a,b}((-q)^{-(a+b)/2})
)(\rc^{a,1}((-q)^{-b-(a+1)/2})\ot I)(s_a^-\ot s^-_{b+1}).\cr}
\nfr{ummm}
The discussion proceeds in the same way.
Assuming \proj\ for $\rc^{a,b}(x)$ the right--hand side of \ummm\ is
$$(I\ot s^-_{a+b})(\rc^{a,1}((-q)^{-b-(a+1)/2})\ot I)(s^-_a\ot s^-_{b+1
}).
\nfr{numi}
By applying \AL\ $a-1$ times we have
$$\rc^{a,1}(x)=\rc_{1}(x(-q)^{(a-1)/2})\rc_{2}(x(-q)^{(a-3)/2})\cdots
\rc_{a}(x(-q)^{-(a-1)/2})s_{1,a}^-.$$
Using Lemma 1 we can rewrite this as
$$s_{2,a+1}^-\rc_{1}(x(-q)^{-(a-1)/2})\rc_{2}(x(-q)^{-(a-3)/2})\cdots
\rc_{a}(x(-q)^{(a-1)/2}).$$
Plugging this into \numi, we have
$$s^-_{2,a+b+1}\rc_{1}((-q)^{-a-b})\rc_{2}((-q)^{-a-b+1})\cdots\rc_{a}(
(-q)^{-b-1})s^-_{a+1,a+b+1}.$$
But by \twof\ this is proportional to
$s_{a+b+1}^-$, as required. This completes the proof.

\appendix{Appendix B}

In this appendix we find the function $f(x)$ that satisfies equations
\BA\ and \BB. By using \BA\ we may rewrite \BB\ as
$$\prod_{i=0}^{n-1}f(x^{-1}(-q)^{-i})=((-q)^{-1}x-(-q)x^{-1})^{-1}\prod_{i=1}^{n-1}(
x^{-1}(-q)^{-i}-x(-q)^i)^{-1}.$$
Now we introduce $u(x)=((-q)^{-1}x^{-1}-(-q)x)f(x)$, where \BA\ implies
$u(x)u(x^{-1})=1$. $u(x)$ satisfies
$$\prod_{i=0}^{n-1}u(x^{-1}(-q)^{-i})=\prod_{i=1}^{n-1}{x(-q)^{i-1}-x^{-1}(-q)^{-i+1
}\over x^{-1}(-q)^{-i}-x(-q)^i}.$$
In order to solve this equation for $u(x)$ we introduce the Gamma
function representation of the sine function. Defining
$q=-\exp-i\pi\lambda$ and $x=\exp-i\pi\mu$, so $\mu=-n\lambda\theta/2\pi
i$, we have
$$x^{-1}(-q)^{-a}-x(-q)^a={2\pi i\over\Gamma(\mu+a\lambda)\Gamma(1
-\mu-a\lambda)}.$$
in terms of this
$$\prod_{i=0}^{n-1}u(x^{-1}(-q)^{-i})=\prod_{i=1}^{n-1}{\Gamma(\mu+i\lambda)\Gamma
(1-\mu-i\lambda)\over\Gamma(-\mu+(1-i)\lambda)\Gamma(1+\mu-(1
-i)\lambda)}.
\nfr{oppp}
To solve this equation we introduce the notation
$$w(x)=\prod_{i=0}^{n-1}\Gamma(x+i\lambda),$$
in terms of which
$$\prod_{i=1}^{n-1}\Gamma(\mu+i\lambda)=\prod_{j=1}^\infty{w(\mu+((j-1)n+1)
\lambda)\over w(\mu+jn\lambda)},$$
and similarly for the other terms in \oppp. Using these facts we can
write down an expression for $u(x)$ as an infinite product of Gamma
functions. In fact because $((-q)^{-1}x^{-1}-(-q)x)^{-1}=\Gamma(\mu+\lambda)
\Gamma(1-\mu-\lambda)/2\pi i$ the explicit expression for $f(x)$ is
$$\eqalign{f(x)=&{\Gamma(\mu+\lambda)\Gamma(1-\mu-\lambda)\over2\pi i}
\prod_{j=1}^\infty{\Gamma(1+\mu+(j-1)n\lambda)\over\Gamma(1-\mu+(j-1
)n\lambda)}\cr &\ \ \ \times{\Gamma(\mu+jn\lambda)\Gamma(-\mu+((j-1)n+1
)\lambda)\Gamma(1-\mu+(jn-1)\lambda)\over \Gamma(-\mu+jn\lambda)\Gamma
(\mu+((j-1)n+1)\lambda)\Gamma(1+\mu+(jn-1)\lambda)}.\cr}
\efr

\references

\beginref
\Rref{jae}{M.T. Jaekel, `{\sl $N$-soliton quantum scattering in the
sine-Gordon theory\/}', Nucl. Phys. {\bf B118} (1977) 506}
\Rref{arin}{A.E. Arinstein, V.A. Fateev and A.B. Zamolodchikov, `{\sl
Quantum $S$-matrix of the $1+1$-dimensional Toda chain\/}', Phys.
Lett. {\bf87B} (1979) 389}
\Rref{hirota}{R. Hirota, `{\sl Direct methods in soliton theory\/}',
in `{\sl Soliton\/}' page 157: ed. R.K. Bullough and P.S. Caudrey
(1980)}
\Rref{wkb}{R.F. Dashen, B. Hasslacher and A. Neveu, `{\sl Particle
spectrum in model field theories from semiclassical functional
integral techniques\/}', Phys. Rev. {\bf D11} (1975) 3424; `{\sl
Nonperturbative methods and extended-hadron models in field theory.
II. Two-dimensional models and extended hadrons\/}', Phys. Rev. {\bf
D10} (1974) 4130}
\Rref{coleman}{S. Coleman, `{\sl The quantum sine-Gordon equation as
the massive Thirring model\/}', Phys. Rev. {\bf D11} (1975)
2088}
\Rref{bpz}{A.A. Belavin, A.M. Polyakov and A.B. Zamolodchikov, `{\sl
Infinite conformal symmetry in two-dimensional quantum field
theory\/}', Nucl. Phys. {\bf B241} (1984) 333}
\Rref{sgle}{D. Bernard and A. LeClair, `{\sl Residual quantum
symmetries of the restricted sine-Gordon theories\/}', Nucl. Phys.
{\bf B340} (1990) 721}
\Rref{sgpert}{A. LeClair, `{\sl Restricted sine-Gordon theory and the
minimal conformal series\/}', Phys. Lett. {\bf230B} (1989) 103\newline
D. Bernard and A. LeClair, `{\sl The fractional supersymmetric sine-Gordon
models\/}', Phys. Lett. {\bf247B} (1990) 309\newline
C. Ahn, D. Bernard and A. LeClair, `{\sl
Fractional supersymmetries in perturbed coset CFTs and integrable
soliton theories\/}', Nucl. Phys. {\bf B346} (1990) 409\newline
F.A. Smirnov, `{\sl Reductions of the sine-Gordon model as perturbations of
conformal field theory\/}', Nucl. Phys. {\bf B337} (1990) 156;
`{\sl The perturbed $c<1$ conformal field theories as reductions of the
sine-Gordon model\/}', Int. J. Mod. Phys. {\bf A4} (1989) 4213\newline
N. Yu Reshetikhin and F. Smirnov, `{\sl Hidden quantum group symmetry
and integrable perturbations of conformal field theories\/}', Comm. Math. Phys.
{\bf131} (1990) 157\newline
T. Eguchi and S-K. Yang, `{\sl Sine-Gordon theory at rational values
of the coupling constant and minimal conformal models\/}', Phys.
Lett. {\bf235B} (1990) 282}
\Rref{zam}{A.B. Zamolodchikov and Al. B. Zamolodchikov, `{\sl
Factorizable $S$-matrices in two dimensions as the exact solutions of
certain relativistic quantum field theories\/}'. Ann. Phys. {\bf120}
(1979) 253\newline Al. B. Zamolodchikov, `{\sl Exact two-particle
$S$-matrix of quantum sine-Gordon solitons\/}'
Comm. Math. Phys. {\bf55} (1977) 183}
\Rref{jimbo}{M. Jimbo, `{\sl A $q$-difference analogue of $U(g)$ and
the Yang-Baxter equation\/}', Lett. Math. Phys. {\bf10} (1985) 63;
`{\sl A $q$-analogue of $U(gl(N+1))$, Hecke algebra, and the
Yang-Baxter equation\/}', Lett. Math. Phys. {\bf11} (1986) 247;
`{\sl Introduction to the Yang-Baxter equation\/}', Int. J. Mod. Phys.
{\bf A4} (1989) 3759}
\Rref{factorize}{A.M. Polyakov, `{\sl Hidden symmetry of the
two-dimensional chiral fields\/}', Phys. Lett. {\bf72B} (1977) 224}
\Rref{Vr}{P. Di Francesco and J.-B Zuber, `{\sl $SU(N)$ lattice integrable
models associated with graphs\/}', Nucl. Phys. {\bf B338} (1990) 602}
\Rref{todasm}{H.W. Braden, E. Corrigan, P.E. Dorey and R. Sasaki, `{\sl
Affine Toda field theory and exact $S$-matrices\/}', Nucl. Phys.
{\bf B338} (1990) 689}
\Rref{qis}{E. Sklyanin, L. Takhtadzhyan and L. Faddeev, `{\sl Quantum
inverse method I\/}', Theor. Math. Phys. {\bf 40} (1980) 688}
\Rref{cardy}{J.L Cardy and G. Mussardo, `{\sl $S$-matrix of the
Yang-Lee edge singularity in two dimensions\/}', Phys. Lett. {\bf225}
(1989) 275}
\Rref{leclair}{D. Bernard and A. LeClair, `{\sl Non-local currents in
2-d QFT: an alternative to the quantum inverse scattering method\/}',
Saclay preprint SACLAY-SPHT-90-173}
\Rref{semiclass}{R. Jackiw and G. Woo, `{\sl Semiclassical scattering
of quantized nonlinear waves\/}', Phys. Rev. {\bf D12} (1975) 1643}
\Rref{korep}{V.E. Korepin, `{\sl Direct calculation of the $S$-matrix
in the massive Thirring model\/}', Theor. Math. Phys. {\bf41}
(1979) 953; `{\sl The mass spectrum and $S$-matrix of the massive
Thirring model in the repulsive case\/}', Comm. Math. Phys. {\bf76} (1980) 165}
\Rref{mysol}{T.J. Hollowood, `{\sl Solitons in affine Toda field
theories\/}', Oxford University preprint OUTP-92-04P}
\Rref{olivet}{D. Olive and N. Turok, `{\sl Local conserved densities
and zero curvature conditions for Toda lattice field theories\/}',
Nucl. Phys. {\bf B257} (1985) 277; `{\sl Algebraic structure of Toda
theories\/}', Nucl. Phys. {\bf B220} (1983) 491}
\Rref{reflect}{V.E. Korepin, `{\sl Above-barrier reflection of
solitons\/}', Theor. Math. Phys. {\bf34} (1978) 1}
\Rref{jon}{J.M. Evans, `{\sl Complex Toda theories and twisted reality
conditions\/}', Oxford University preprint OUTP-91-39P}
\Rref{tim}{T.J. Hollowood, `{\sl A quantum group approach to
constructing factorizable $S$-matrices\/}', Oxford University preprint
OUTP-90-15P}
\endref
\ciao